\documentclass[12pt,fleqn]{article}
\usepackage{amsfonts,ulem,url}
\usepackage{amssymb}
\usepackage{amsmath}
\usepackage{amstext,verbatim,graphicx,ifthen,multirow,amsthm}
\usepackage{bm}
\usepackage{natbib}
\usepackage[bf]{caption}
\usepackage{subcaption}
\usepackage{setspace}
\usepackage[pdftex,colorlinks=true,unicode,bookmarksnumbered=false, hyperfootnotes=true]{hyperref}
\hypersetup{citecolor = blue}

\usepackage[margin=1in]{geometry}

\usepackage{tkz-graph}
\usetikzlibrary{positioning}
\usetikzlibrary{arrows.meta}

\tikzstyle{observed}=[circle, inner sep=0mm, outer sep=0mm, minimum size=2mm, draw=black, fill=black]
\tikzstyle{unobserved}=[circle, inner sep=0mm, outer sep=0mm, minimum size=2mm, draw=black, fill=white]

\tikzstyle{notouch}=[shorten <=5pt, shorten >= 5pt, -{Latex[length=2mm, width=1.5mm]}]

\makeatletter
\g@addto@macro\@floatboxreset\centering
\makeatother

\newcommand{\obs}{{\rm obs}}
\newcommand{\bd}{{\rm bd}}
\newcommand{\indep}{\perp\!\!\!\perp}

\newcommand{\doo}{{\it do}}
\newcommand{\mmz}{\mathbb{Z}}

\def\monthname{\ifcase\month\or
  January\or February\or March\or April\or May\or June\or July\or
  August\or September\or October\or November\or December\fi}
\numberwithin{equation}{section}

\def\monthname{\ifcase\month\or
January\or February\or March\or April\or May\or June\or
July\or August\or September\or October\or November\or December\fi}
\renewcommand{\appendix}{\small\parindent 0cm\setcounter{equation}{0}
\renewcommand{\theequation}{A.\arabic{equation}}
\setcounter{lemma}{0}\renewcommand{\thelemma}{A.\arabic{lemma}}
\setcounter{theorem}{0}\renewcommand{\thetheorem}{A.\arabic{theorem}}}
\begin{document}

\title{\textbf{Potential Outcome 
and Directed Acyclic Graph Approaches
to Causality: Relevance for Empirical Practice in Economics}\thanks{\small I am
grateful for help with the graphs by Michael Pollmann and for comments by  Alberto Abadie, Jason Abaluck,  Alexei Alexandrov,
Josh Angrist, Susan Athey, Gary Chamberlain, Stephen Chaudoin,
 Rebecca Diamond, Dean Eckles, Ernst Fehr, Avi Feller, Paul Goldsmith-Pinkham, Chuck Manski, Paul Milgrom, Evan Munro, Franco Perrachi,  Michael Pollmann, Thomas Richardson, Jasjeet Sekhon, Samuel Skoda, Korinayo Thompson,
and Heidi Williams. They are not responsible for any of the views expressed here.
Financial support from the Office of Naval Research under grant N00014-17-1-2131 is gratefully acknowledged. The first version of this paperwas  circulated in May 2019.}} 
\author{Guido W. Imbens\thanks{\small Professor of
Economics,
Graduate School of Business, and Department of Economics, Stanford University, SIEPR, and NBER,
imbens@stanford.edu.} }
\date{\ifcase\month\or
January\or February\or March\or April\or May\or June\or
July\or August\or September\or October\or November\or December\fi \ \number%
\year\ \ }

\maketitle\thispagestyle{empty}

\begin{abstract}
In this essay I discuss potential outcome and graphical approaches to causality, and their relevance for empirical work in economics. I review some of the work on directed acyclic graphs, including the recent ``The Book of Why,'' (\citep*{pearl2018book}). I also discuss the potential outcome framework developed by Rubin and coauthors ({\it e.g.,} \citep*{rubin2006matched}), building on work by Neyman (\citep{neyman1923}). I then discuss  the relative merits of these approaches for empirical work in economics, focusing on the questions each framework answer well, and why much of the  the work in economics is closer in spirit to the potential outcome perspective.
\end{abstract}

\begin{center}
\end{center}



\baselineskip=20pt\newpage
\setcounter{page}{1}
\renewcommand{\thepage}{\arabic{page}}
\renewcommand{\theequation}{\arabic{section}.\arabic{equation}}

\section{Introduction}

Causal Inference (CI) in observational studies has been an integral  part of econometrics since its start as a separate field in the 1920s and 1930s. 
The simultaneous equations methods developed by \citep*{tinbergen1930determination}, \citep*{wright1928tariff},   \citep*{haavelmo1943statistical}, and their successors in the context of supply and demand settings were from the beginning, and continue to be, explicitly focused on causal questions. 
Subsequently, the work by the Cowles commission, and both the structural and reduced form approaches since then, have thrived by focusing on identifying and estimating causal and policy-relevant parameters.
Over the last thirty years close connections to related research on causal inference  in other social sciences and statistics, and, more recently,  computer science have been established.  In this essay I review some of the approaches to causal inference  in the different disciplines in the context of some recent textbooks, and 
discuss the relevance of these perspectives for empirical work in economics. The two main frameworks are $(i)$ the {\it Potential Outcome} (PO) framework, associated with the work by Donald Rubin, building on the work on randomized controlled trials (RCTs)  from the 1920s by Ronald Fisher and Jerzey Neyman, and $(ii)$ the work on {\it Directed Acyclic Graphs} (DAGs), much of it associated with work by Judea Pearl and his collaborators.
These frameworks are complementary, with different strengths that make them particularly appropriate for different questions. Both have antecedents in the earlier econometric literature, the PO framework in the demand and supply models in \citep*{tinbergen1930determination}  and  \citep*{haavelmo1943statistical}, and the DAG approach in the path analysis in  \citep*{wright1928tariff}.

The body of  this essay consists of three parts. In the first part of the essay I  discuss the graphical approach to causality in the context of the recent book
``The Book of Why'' (TBOW) by 
Judea Pearl and Dana Mackenzie 
\citep*{pearl2018book}, which is a very accessible  introduction to  this approach.
 I highly recommend both 	TBOW and the more technical and comprehensive \citep*{pearl}   for anyone who is interested in causal inference in statistics, and that should include pretty much anyone doing empirical work in social sciences these days.
The graphical approach has gained much traction in the wider  computer science community, see for example 
the recent text ``Elements of Causal Inference,'' (\citep*{peters2017elements}) and the work on causal discovery (\citep*{glymour2014discovering, hoyer2009nonlinear, lopez2017discovering, ke2019learning}) and interpretability (citet{zhao2019causal}), and also in parts of epidemiology and social sciences, although it has not had as much impact in economics as it should have
Exceptions include the methodological discussions in
\citet{white2006unified, white2009settable, chalak2011extended, heckman2015causal}, and the recent  discussions on on the benefits  graphical causal modeling for economics in
\citet{mixtape, hunermund2019causal}.

  In the second part of this essay, in Section \ref{section:po}, I   review the potential outcome approach, associated with the work by Donald Rubin, Paul Rosenbaum, and collaborators, that is more familiar to economists. Representative of the methodological part of the potential outcome literature is
the collection of papers by Rubin and coauthors,  ``Matched Sampling for Causal Effects,'' (MACE, \citep*{rubin2006matched}) and  ``Observation and Experiment,'' (\citep*{rosenbaum2017observation}).  Other   references to this literature include \citep*{rubin1974estimating, rosenbaum_book, rosenbaum2010design},  \citep*{holland1986statistics} which coined the term ``Rubin Causal Model''  for this approach, and my own text with Rubin, ``Causal Inference in Statistics, Social, and Biomedical Sciences,'' (CISSB, \citep*{imbens2015causal}).\footnote{As a disclaimer, I have worked with both Rubin and Rosenbaum, and my own work on causal inference is largely within the PO framework, although some of that work preceeds these collaborations.}

 In the third and main part of this essay, in Section \ref{section:practice},  I   discuss 
 the comparative strengths and weaknesses    of the PO and DAG approaches. I  discuss  why the graphical approach to causality has not caught on  very much (yet) in economics. For example, a recent econometrics textbook  focused on causal inference,  Mostly Harmless Econometrics (MHE, \citep*{angrist2008mostly}), has no DAGs, and is largely in the PO spirit.
 Why has it not caught on, or at least not yet?
 
At a high level the DAG approach has two fundamental benefits to offer. First, in what is essentially an pedagogical component, the formulation of the critical assumptions is intended to capture the way researchers think of causal relationships. 
 The DAGs, like the path analyses that came before them (\citep*{wright1928tariff, wright1934method}), can be a powerful  way of illustrating key assumptions in causal models.
 I  elaborate on this aspect of DAGs in the discussions of instrumental variables and mediation/surrogates. 
	 Ultimately some of this  is a matter of taste and some researchers may prefer graphical versions to algebraic versions of the same assumptions and {\it vice versa}.
Second, the machinery developed in the DAG literature, in particular the \doo-calculus discussed in Section \ref{doo}, aims to allow researchers to  answer particular causal queries in a systematic way.
Here the two frameworks are complementary and have different strengths and weaknesses. The DAG machinery  simplifies the answering of certain causal queries compared to the PO framework. This is particularly true for queries in complex models with a large number of variables. 
However, there are many causal questions in economics for which this is is not true.

In comparison, there are five features of the PO framework that  may be behind its current  popularity in economics.  
First, there are some assumptions that are easily captured in the PO framework relative to the DAG approach, and these assumptions are critical in many identification strategies in economics. Such assumptions include  monotonicity (\citep*{imbens1994})  and other shape restrictions such as  convexity or concavity (\citep*{matzkin1991semiparametric, chetverikov2018econometrics, chen2018shape}). The instrumental variables setting is a prominent example where such shape restrictions are important, and I will discuss it in  detail in Section \ref{section:iv}.
Second, the potential outcomes in the PO framework connect easily to traditional  approaches to economic models such as supply and demand settings where potential outcome functions are the natural primitives.
 Related to this, the insistence of the PO approach on manipulability of the causes, and its attendant distinction between non-causal attributes  and causal variables has resonated well with the focus in empirical work on policy relevance (\citep*{angrist2008mostly, manski2013public}). Third, many of the currently popular identification strategies focus on   models with relatively few (sets of) variables, where identification questions have been worked out once and for all. 
 Fourth, the PO framework lends itself well to accounting for  treatment effect heterogeneity in estimands (\citep*{imbens1994, sekhon2017inference}) and incorporating such heterogeneity in estimation and the design of optimal policy functions (\citep*{athey2017efficient, athey2019generalized, kitagawa2015should}).
 Fifth, the PO approach has traditionally connected well with questions of study design, estimation of causal effects and inference for such effects. From the outset Rubin and his coauthors provided much guidance to researchers and policy makers for practical implementation including inference, with the work on the propensity score (\citep*{rosenbaum1983central}) an influential example.

Separate from the theoretical merits of the two approaches, another  reason for the lack of adoption in economics is that the DAG literature  has not shown much evidence of the alleged benefits for empirical practice in settings that resonate with economists. The potential outcome studies in MACE, and the chapters in \citep*{rosenbaum2017observation}, CISSB and MHE have detailed empirical examples of the various identification strategies proposed. In  realistic settings they demonstrate the merits of the proposed methods  and describe in detail the corresponding estimation and inference methods. In contrast  in the DAG literature,  TBOW, \citep*{pearl}, and \citep*{peters2017elements} have  no substantive empirical examples, focusing largely on identification questions in  what TBOW refers to as { ``toy''} models. 
Compare the lack of impact of the DAG literature in economics with  the recent embrace of regression discontinuity designs imported from the psychology literature, or with the current rapid spread of the machine learning methods from computer science, or the recent quick adoption of synthetic control methods developed in economics \citep*{abadie2003, abadie2010synthetic}. All three came with multiple concrete  and detailed examples that highlighted their benefits over traditional methods. 
In the absence of such concrete examples the toy models in the DAG literature  sometimes appear to be a set of solutions in search of problems, rather than a set of clever solutions for substantive problems previously posed in social sciences, bringing to mind the discussion of Leamer on the Tobit model (\citep{leamer1997revisiting}).


\section{The Graphical Approach to Causality and TBOW}\label{section:dags}

In this section I review parts of TBOW and give a brief general introduction to Directed Acyclic Graphs (DAGs).

\subsection{TBOWs View of Causality, and the Questions of Interest}\label{section:2.2}

Let me start by clarifying what TBOW, and in general the DAG approach to causality is interested in.
The primary focus of  TBOW, as well as \citep*{pearl}, is  on {\it identification}, as opposed to estimation and inference.  As 
Figure 1 (TBOW, p. 12) illustrates, researchers arrive armed with a number of variables and a causal model linking these variables, some observed and some unobserved. 
The assumptions underlying this model are coded up in a graphical model, a Directed Acyclic Graph, or DAG.
The researchers then  ask causal queries.
Early in TBOW the authors present some examples of such questions (TBOW, p. 2):
\begin{enumerate}
\item How effective is a given treatment in preventing a disease?
\item Did the new tax law cause our sales to go up, or was it our advertising campaign?
\item What is the health-care cost attributable to obesity?
\item Can hiring records prove an employer is guilty of a policy of sex discrimination?
\item I'm about to quit my job. Should I?
\end{enumerate}
These types of questions are obviously all of great importance. Does the book deliver on this, or more precisely, does the methodology described in the book allow us to answer them? The answer essentially is an indirect  one: if you tell me how the world works (by giving me the full causal graph), I can tell you the answers. Whether this is satisfactory  really revolves around how much the researcher is willing to assume about how the world works. Do I feel after reading the book that I understand better how to answer these questions? That is not really very clear.
The rhetorical device of giving these specific examples at the beginning of the book is very helpful, but  the book does not really provide context for them. Questions of this type have 
 come up many times before, but there is little discussion of the previous approaches to answer them. The reader is never given the opportunity to compare previously proposed answers.
 It would have been helpful for the reader if in the final chapter TBOW would have returned to these five questions and described specific answers given the book to these five questions and then compared them to alternative approaches. Instead some of the questions come back at various stages, but not systematically, and the impression is created that it was not possible to answer these questions previously.

One class of questions that is missing from this list, somewhat curiously given the title of TBOW,  is explicitly ``why'' questions. 
Why did Lehmann Brothers collapse in 2008? Why did the price of a stock go up last year? Why did unemployment go down in the Great Depression?
 \citep*{gelman2013ask} refer to such questions as {\it reverse} causal inference question, ``why'' an outcome happened, rather than {\it forward} causal questions that are concerned with the effect of a particular manipulation. The causal literature in general has focused much less on such reverse causal questions.

The focus of TBOW, \citep*{pearl}  and \citep*{peters2017elements},  is on developing machinery for  answering  questions  
of this type
given two inputs. First, knowledge of  the joint distribution of all the observed  variables in the model, and, second, a causal model describing the phenonoma under consideration. 
Little is said about what comes before the identification question, namely the development of the model, and what comes after the identification question, namely estimation and inference given a finite sample.
 The position appears to be that the specification of a causal model and the statistical analyses are  problems separate from those of identification, with the integration of those problems with questions of identification less important than the benefits of the specialization associated with keeping the identification questions separate.

However, many  statistical problems and methods are specific to the  causal nature of the questions, and as a result much of the methodological literature on causality in statistics and econometrics is about estimation methods. This includes the literature on weak instruments \citep*{stock1997, andrews2007}, the literature on unconfoundedness including discussions of the role of the propensity score (\citep*{rosenbaum1983central}) and problems with overlap (\citep*{crump2009dealing, d2017overlap, li2018balancing}), double robustness \citep*{robins1, imbens2004,  belloni2013program, athey2018approximate},  the literature on regression discontinuity designs \citep*{hahntodd, imbenskalyanaraman}, and the recent work on estimating heterogenous treatment effects \citep*{athey2016recursive, wager2017estimation} and synthetic control methods (\citep*{abadie2003, abadie2010synthetic}.
Another area where the separation between identification of causal effects and the identification of the joint distribution of realized variables is more difficult is in network settings \citep*{graham2015methods, ogburn2014causal, athey2018exact}. This integration of statistical questions and causal identification has arguably been very beneficial in many of these settings.

The choices and challenges in postulating a causal model, graphical or otherwise, that is, a model of how the world works, is also not a major subject of the book. TBOW views that as the task of subject matter experts:
\begin{quote}
``I am not a cancer specialist, and I would always have to defer to the expert opinion on whether such a diagram represents and real-world processes accurately.'' (TBOW, p. 228)\end{quote}
and
\begin{quote}``I am not personally an expert on climate change`' (TBOW, p. 294)
\end{quote}
This is of course fine,  but the result is that the models in TBOW are all, in a favorite phrase of Pearl's, ``toy models,'' suggesting that we should not take them too seriously. This is common to other discussions of graphical causal models ({\it e.g.}, \citep*{koller2009probabilistic}). It would have been useful  if the authors had teamed up with subject matter experts and discussed some substantive examples where DAGs, beyond the very simple ones implied by randomized experiments, are  aligned with experts' opinions. Such examples, whether in social sciences or otherwise, would serve well in the effort to convince economists that these methods are useful in practice.

The focus on toy models and the corresponding lack of engagement with estimation and inference questions is  in sharp contrast to the econometrics literature  where the three steps, $(i)$ the development of the causal models that preceeds the identification question,  $(ii)$ the study of identification questions, and $(iii)$ estimation and inference methods that follow once the identification questions have been resolved, typically go hand-in-hand. 
The models in econometric papers are often developed with the idea that they are useful on settings beyond the specific application in the original paper. 
Partly as a result of the focus on empirical examples the econometrics literature has developed a small number of canonical settings where researchers view the specific causal models 
and associated statistical methods
as well-established and understood. 
These causal models  correspond to what is nowadays often referred to as {\it  identification strategies} ({\it e.g.,} \citep*{card1993using, angristkruegerstrategies}).
These identification strategies that include adjustment/unconfoundedness, instrumental variables, difference-in-differences, regression discontinuity designs, synthetic control methods (the first four are listed in ``Empirical Strategies in Labor Economics,'' \citep*{angristkruegerstrategies})
are widely taught in both undergraduate and graduate economics courses, and they are familiar to most empirical researchers in economics. The statistical methods associated with 
these causal models are commonly used in empirical work and are constantly being refined, and new identification strategies are occasionally added to the canon. 
Empirical strategies not currently in this canon, rightly or wrongly, are  viewed with much more suspicion until they reach the critical momentum to be included. This canon is not static: despite having been developed in the early 1960s in the psychology literature, regression discontinuity designs were virtually unheard of in economics until the early 2000s when a number of examples caught researchers' attention  ({\it e.g.,} \citep*{black1999better, vanderklaauw, leemoretti, pettersson2008parties}). Now  regression discontinuity designs are commonly taught in  graduate and undergraduate courses. Similarly synthetic control
methods (\citep*{abadie2010synthetic}) have become very popular in a very short period of time, and are a staple of graduate econometrics courses.

\subsection{The Ladder of Causality}

TBOW introduces a classification of causal problems that they call the {\it Ladder of Causality,} with three rungs, in order of complexity labeled association, intervention, and counterfactuals respectively. 

On the first rung, assocation, researchers observe passively, and form predictions based on these observations. A key concept is that of {\it correlation} or {\it association}. Methods belonging to this rung according to the discussion in TBOW are regression, as well as many of the modern machine learning methods such as regression trees, random forests, and deep neural nets. Of course regression is used in many disciplines as a causal method, but here TBOW views regression in something akin to what econometricians would call the best linear predictor framework, where the regression function is simply a parametric way of fitting the conditional expectation \citep*{goldberger1991course}. There is little causal in this rung, and the problems here are well understood and continue to be  studied in a variety of disciplines. They are routinely taught in economics PhD programs as part of the econometrics or statistics curriculum. Much of this is now being integrated with predictive machine learning methods (\citep*{atheyimbens2018}).

The second rung is that of intervention. A canonical example, used in Figure 1.2 (TBOW, p. 28), and also in CISSB (p. 3),  is the question what would happen to my headache if I take an aspirin. In general the questions in this rung are about manipulations. These are the questions that much of the causal inference work in the PO framework  is focused on. Randomized experiments are one of the key statistical designs here.  In observational studies these questions are much harder, but they are studied in a wide range of areas using a wide range of methods.  Question 1 in the list of questions in TBOW (``How effective is a given treatment in preventing a disease?'') belongs on this rung. This is where much of the empirical work in economics takes place. The challenges typically concern the presence of unobserved confounders of some type or another because economist typically model the behavior of optimizing agents, who often are more knowledgable than the researcher and who take into account the expected results of their actions. 
The identification strategies in \citep*{angristkruegerstrategies} fit in here.

The third rung of the ladder of causality deals with counterfactuals. Here the type of question considered is ``What would have happened had  I not taken the aspirin?'' [given that I did take the aspirin, GWI] (TBOW, p. 33). 
The questions on this third run are more difficult to answer, and the PO framework is more apprehensive about definite answers to such questions that depend delicately on individual-level heterogeneity. In the PO framework the correlation between the potential outcomes given the aspirin and without the aspirin, within subpopulations homogenous in observed characteristics, is not point-identified, As a result estimands that depend on this correlation, which includes most questions on the third rung, are only partially identified. 
Although in legal settings this type of question does come up routinely in so-called ``but-for'' analyses,\footnote{But for the existence of X, would Y have occurred?} the economics literature does not focus as much on this type of question as it does on the second type.

Related to the issue raised in the discussion in Section \ref{section:2.2} of the list of questions provided in TBOW, I would have liked to have seen a fourth rung of the ladder, dealing with ``why,'' or reverse causality questions (\citep*{gelman2013ask}). Such questions are related to both the second and third rung, but not quite the same.

\subsection{Directed Acyclic Graphs}

The approach to causality in TBOW and \citep*{pearl}  centers on graphical models, and in particular {Directed Acyclic Graphs} (DAGs). 
These are seen as an attractive way to  capture how people think about causal relationships.
The DAGs are characterized by {\it nodes} and {\it directed edges} between the nodes. Let us start with four  examples, in increasing order of complexity.

\begin{figure}
    \centering
    \begin{subfigure}[b]{0.45\textwidth}  
    \begin{tikzpicture}[
        >=stealth,
        node distance=1.5cm        ]
        \node[observed, label=above:{\(X\)}] (1) {};
        \node[observed, right=of 1,  label=above:{\(Y\)}] (2) {};
        \draw [solid, ->, notouch]  (1.west) -- (2.east);
    \end{tikzpicture}
            \caption{\label{rd_a} Randomized Experiment}
    \end{subfigure}
    \hfill
    \begin{subfigure}[b]{0.45\textwidth}   
    \vspace{2em}
    \begin{tikzpicture}[
        >=stealth,
        node distance=1.5cm
        ]
        \node[observed, label=above:{\(X\)}] (1) {};
        \node[observed, right=of 1,  label=above:{\(Y\)}] (2) {};
        \draw [->, notouch]   (2.east) -- (1.west);
    \end{tikzpicture}
        \caption{\label{rd_b} Reverse Causality}
    \end{subfigure}
    \vspace{3em}
    \begin{subfigure}[b]{0.45\textwidth}
        \vspace{2em}
    \begin{tikzpicture}\label{unobs}[
        >=stealth,
        node distance=1.5cm
        ]
        \node[observed, label=above:{\(X\)}] (1) {};
        \node[observed, right=of 1,  label=above:{\(Y\)}] (2) {};
        \node[unobserved, below right=of 1, label=below:{\(U\)}] (3) {};
        \draw [dashed, ->, notouch] (3.north west) -- (1.south east);
        \draw [dashed, ->, notouch] (3.north west) -- (2.south west);
    \end{tikzpicture}    
        \caption{\label{rd_c}  Spurious Correlation}
    \end{subfigure}
    \hfill
    \begin{subfigure}[b]{0.45\textwidth}\vspace{2em}
    \begin{tikzpicture}[
        >=stealth,
        node distance=1.5cm
        ]
        \node[observed, label=above:{\(X\)}] (1) {};
        \node[observed, right=of 1,  label=above:{\(Y\)}] (2) {};
        \node[unobserved, below=of 2, label=below:{\(U_2\)}] (3) {};
     \node[unobserved, below=of 1, label=below:{\(U_1\)}] (4) {};
        \draw [->, notouch] (1.east) -- (2.west);
        \draw [dashed, ->, notouch] (4.north) -- (1.south east);
        \draw [dashed, ->, notouch] (3.north) -- (2.south west);
    \end{tikzpicture}
        \caption{\label{rd_d}  Randomized Experiment (Alternative DAG)}
    \end{subfigure}
  \caption{\label{fig_randomized_experiment} DAGs for the Two Variable Case}
\end{figure}
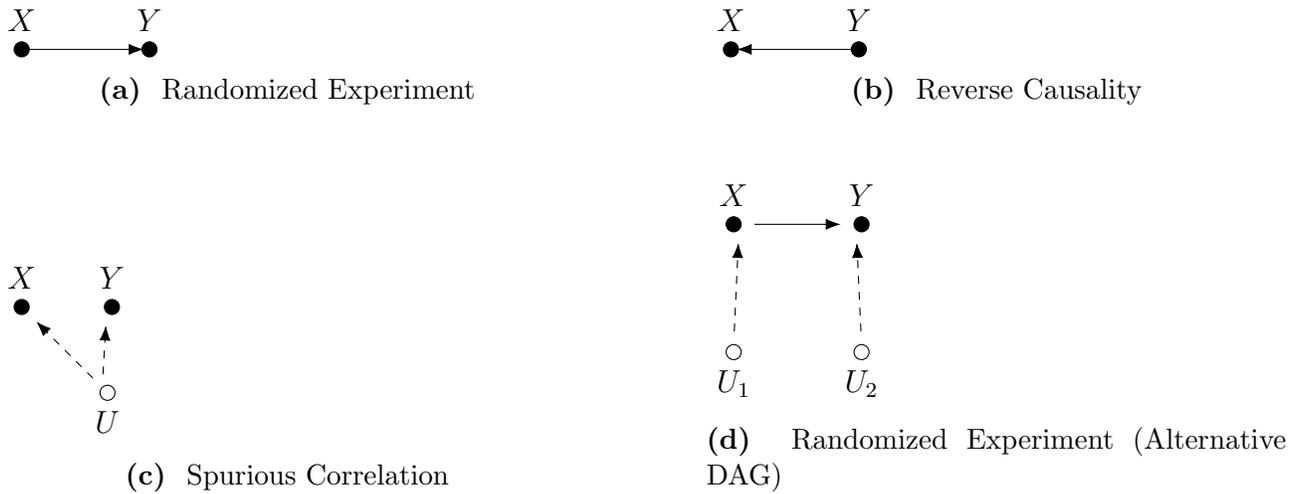


The first example is as simple as it gets. In Figure \ref{fig_randomized_experiment}(\subref{rd_a}) there are only two nodes, corresponding to two variables, denoted by $X$ and  $Y$. There is a single arrow (directed edge) connecting these two nodes, going from $X$ to $Y$. The direction of the arrow captures the notion that $X$ ``causes'' $Y$, rather than some other explanation.
Alternative explanations include  $Y$ causing $X$ as in Figure  \ref{fig_randomized_experiment}(\subref{rd_b}), or some third, unobserved, variable $U$ (denoted by a circle rather than a solid dot to denote that it is not observed) causing both, as in the spurious correlation in Figure  \ref{fig_randomized_experiment}(\subref{rd_c}). 
If we have data on the values of these variables $X$ and $Y$ for a large number of units (meaning we can infer  the full joint distribution of $X$ and $Y$), we can estimate the association between them. The model then allows us to infer from that association  the causal effect of $X$ on $Y$. Obviously simply the data on $X$ and $Y$ are not sufficient: we need the causal model to go from the association to the causal statement that it is $X$ causing $Y$ and not the other way around. The model also says more than simply coding the direction of the causal link. It also captures, through the absence of other nodes and edges, the fact that there are no other variables that have causal effects on both $X$ and $Y$ as, for example, in Figure  \ref{fig_randomized_experiment}(\subref{rd_c}).
Note that we could expand the DAG by including two unobserved variables, $U_1$ and $U_2$, with an arrow $(U_1\rightarrow X)$ and an arrow $(U_2\rightarrow Y)$, as in Figure  \ref{fig_randomized_experiment}(\subref{rd_d}). 
In the Structural Equation Model (SEM) version of the DAGs these unobserved variables would be explicit.
Because there is no association between $U_1$ and $U_2$, the presence of these two unobserved variables does not affect any conclusions, so we omit them from the DAG, following convention. 

\begin{figure}
    \vspace{0.5em}
    \begin{tikzpicture}[
        >=stealth,
        node distance=2.0cm
        ]
        \node[observed, label=above:{\(X\)}] (1) {};
        \node[observed, right=of 1,  label=above:{\(Y\)}] (2) {};
        \node[observed, below right=of 1, label=below:{\(W\)}] (3) {};
        \draw [->, notouch] (1.east) -- (2.west);
        \draw [->, notouch] (3.north west) -- (1.south east);
        \draw [->, notouch] (3.north west) -- (2.south west);
    \end{tikzpicture}\vspace{3em}    
    \caption{\label{fig_unc}Unconfoundedness}
   \end{figure}
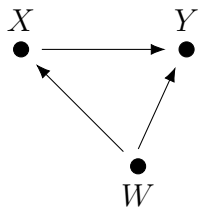

Next, consider Figure \ref{fig_unc}. Here the DAG is slightly more complex. There are now three observed variables. In addition to $X$ and $Y$, with an arrow going from $X$ to $Y$, there is a third variable $W$ with arrows going both from $W$ to $X$ and from $W$ to $Y$. $W$ here is a {\it confounder}, or, more precisely, an {\it observed confounder}. Simply looking at the association between $X$ and $Y$ is not sufficient for infering the causal effect: the effect is confounded by the effect of $W$ on $X$ and $Y$. Nevertheless, because we observe the confounder $W$ we can still infer the causal effect of $X$ on $Y$ by {\it controlling} or {\it adjusting} for $W$.

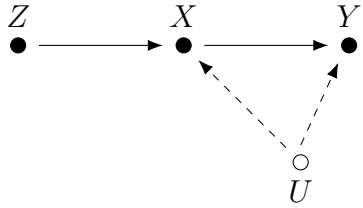
\begin{figure}
    \vspace{2em}
    \begin{tikzpicture}[
        >=stealth,
        node distance=2.0cm
        ]
        \node[observed, label=above:{\(X\)}] (1) {};
        \node[observed, right=of 1,  label=above:{\(Y\)}] (2) {};
        \node[unobserved, below right=of 1, label=below:{\(U\)}] (3) {};
        \node[observed, left=of 1, label=above:{\(Z\)}] (4) {};
        \draw [->, notouch] (1.east) -- (2.west);
        \draw [dashed, ->, notouch] (3.north west) -- (1.south east);
        \draw [dashed, ->, notouch] (3.north west) -- (2.south west);
        \draw [->, notouch] (4.east) -- (1.west);
    \end{tikzpicture}\vspace{3em}
    \caption{\label{fig_iv}Instrumental Variables}
   \end{figure}

Figure \ref{fig_iv} is even more complex. Now there are three observed variables, $X$, $Y$, and $Z$, and one unobserved variable $U$ (denoted by a circle rather than a dot). 
There are arrows from $Z$ to $X$, from $X$ to $Y$, and from $U$ to both $X$ and $Y$. The latter two are dashed lines in the figure to indicate they are between two nodes at least one of which is an unobserved variable.
$U$ is an unobserved confounder. The presence of $U$  makes it in general impossible to completely infer the average causal effect of $X$ on $Y$ from just  the joint distribution of $X$ and $Y$, although bounds as in \citep{manski_bounds}. The presence of the additional variable, the instrument $Z$, is one way  to make progress.
This DAG captures an {\it instrumental variables} setting. 
In the econometrics terminology $X$ is {\it endogenous} because there is an unobserved confounder $U$ that affects both $X$ and $Y$.
There is no direct effect of the instrument $Z$ on the outcome $Y$, and there is no unobserved confounder for the effect of the instrument on the endogenous regressor or the outcome.
This instrumental variables set up is familiar to economists, although traditionally in a  non-DAG form.
In support of his first argument of the benefits of DAGs, TBOW argues that the DAG version clarifies the key assumptions and structure compared to the econometrics set up. Compared to the traditional econometrics setup where the critical assumptions are expressed in terms of the correlation between residuals and instruments, I  agree with TBOW that the DAGs are superior in clarity. I am less convinced of the benefits of the DAG relative to the modern PO set up for the IV setting with its separation of the critical assumptions into design-based unconfoundedness assumptions and a substantive exclusion restriction ({\it e.g.,} \citep*{angrist1996}), but that appears to me to be a matter of taste.
Certainly for many people the DAGs are an effective expository tool. That is quite separate from its value as formal method for infering identification and lack thereof.
Note that  formally in this instrumental variables setting   identification of the average causal effect of $X$ on $Y$ is a subtle one, and in fact this average effect is not identified according to the DAG methodology.  I will return to this setting where only  the Local Average Treatment Effect (LATE) is identified  (\citep*{imbens1994})   in more detail in Section {\ref{section:iv}}. It is an important example because it shows explicitly the inability of the DAGs to derive some classes of identification results.

\begin{figure}
    \centering
    \begin{subfigure}[b]{0.45\textwidth}
        \begin{tikzpicture}[
            >=stealth,
            node distance=1.5cm
            ]
            \node[unobserved, label=above:{\(Z_0\)}] (1) {};
            \node[observed, below=of 1, label=right:{\(Z_1\)}] (3) {};
            \node[unobserved, right=of 3, label=above right:{\(B\)}] (2) {};
            \node[observed, below=of 3, label=below right:{\(Z_2\)}] (5) {};
            \node[observed, left=of 5, label=left:{\(X\)}] (4) {};
            \node[observed, right=of 5, label=right:{\(Z_3\)}] (6) {};
            \node[observed, below=of 5, label=below:{\(Y\)}] (7) {};
            \draw [dashed, ->, notouch] (1.south east) -- (2.north west);
            \draw [dashed, ->, notouch] (1.south) -- (3.north);
            \draw [dashed, ->, notouch] (1.south west) -- (4.north west);
            \draw [dashed, ->, notouch] (2.south) -- (6.north);
            \draw [->, notouch] (3.south) -- (5.north);
            \draw [->, notouch] (4.east) -- (5.west);
            \draw [->, notouch] (5.east) -- (6.west);
            \draw [->, notouch] (4.south) -- (7.north west);
            \draw [->, notouch] (5.south) -- (7.north);
            \draw [->, notouch] (6.south) -- (7.north east);
        \end{tikzpicture}
        \caption{original}
    \end{subfigure}
    \hfill
    \begin{subfigure}[b]{0.45\textwidth}
        \begin{tikzpicture}[
            >=stealth,
            node distance=1.5cm
            ]
            \node[unobserved, label=above:{\(Z_0\)}] (1) {};
            \node[observed, below=of 1, label=right:{\(Z_1\)}] (3) {};
            \node[unobserved, right=of 3, label=above right:{\(B\)}] (2) {};
            \node[observed, below=of 3, label=below right:{\(Z_2\)}] (5) {};
            \node[observed, left=of 5, label=left:{\(X\)}] (4) {};
            \node[observed, right=of 5, label=right:{\(Z_3\)}] (6) {};
            \node[observed, below=of 5, label=below:{\(Y\)}] (7) {};
            \draw [dashed, ->, notouch] (1.south east) -- (2.north west);
            \draw [dashed, ->, notouch] (1.south) -- (3.north);
            \draw [dashed, ->, notouch] (1.south west) -- (4.north west);
            \draw [dashed, ->, notouch] (2.south) -- (6.north);
            \draw [->, notouch] (3.south) -- (5.north);
            \draw [->, notouch] (4.east) -- (5.west);
            \draw [->, notouch] (5.east) -- (6.west);
            \draw [->, notouch] (4.south) -- (7.north west);
            \draw [->, notouch] (5.south) -- (7.north);
            \draw [->, notouch] (6.south) -- (7.north east);
            \draw [dashed, ->, notouch] (2.south east) to [out=-35, in=0,looseness=1.7] (7.east);
            \draw [dashed, ->, notouch] (4.north east) -- (2.south west);
        \end{tikzpicture}
        \caption{Two Additional Links}
    \end{subfigure}
    \caption{\label{fig_pearl} Based on Figure 1 in \citep*{pearl1995causal}.}
\end{figure}
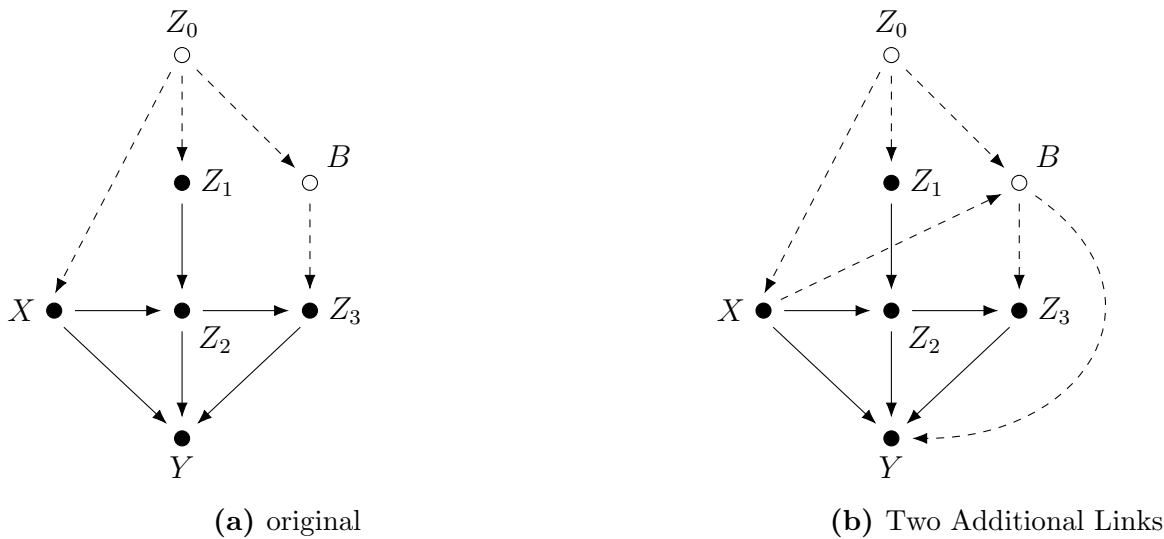

The last example, in Figure \ref{fig_pearl}, is substantially more complex. The first subfigure, Figure  \ref{fig_pearl}(a),  is taken from 
\citep*{pearl1995causal}.
There are five observed variables, 
 soil fumigation ($X$), crop yield ($Y$)
the eelworm population before the treatment ($Z_1$), the eelworm population after the fumigation, ($Z_2$), and the eelworm population at the end of the season, ($Z_3$). There are two unobserved variables, the bird population ($B$) and the eelworm population last season, ($Z_0$). The question is whether  we can identify the effect of the soil fumigation $X$ on the crop yield $Y$ from the joint distribution of the observed variables $(X,Y,Z_1,Z_2,Z_3)$. The \doo-calculus, described in Section \ref{doo} is helpful here.
This is an example of the second benefit of DAGs for causal modeling, the ability to infer identifiability given a complex model. \citep*{pearl} argues that  in contrast to the DAG approach the PO framework is not well equipped to assess identifiability in  complex models involving a large number of variables: 
\begin{quote}``no mortal can apply this condition [ignorability, GWI] to judge whether it holds even in simple problems, with all causal relationships correctly specified, let alone in partially specified problems that involve dozens of variables.'' (\citep*{pearl}, p. 350).
\end{quote}
Similarly, Elias  Bareinboim writes,
\begin{quote}
``Regarding the frontdoor or napkin, these are just toy examples where the current language [the PO framework, GWI] has a hard time solving, I personally never saw a natural solution. If this happens with 3, 4 var examples, how could someone solve one 100-var instance compounded with other issues?'' (Elias Bareinboim, Twitter,@eliasbareinboim,  March 17, 2019).
\end{quote}
I agree that in cases like this inferring identifiability is a challenge that few economists would be equipped  to meet. 
However,  in modern empirical work in economics  there are few cases where researchers  consider models with dozens, let alone a hundred, variables and complex relations between them that do not reduce to simple identification strategies.
As Jason Abaluck responds to Bareinboim's comment:
\begin{quote}
``No one should ever write down a 100 variable DAG and do inference based on that. That would be an insane approach because the analysis would be totally impenetrable. Develop a research design where that 100 variable DAG trivially reduces to a familiar problem (e.g. IV!)''
(Jason Abaluck, Twitter, @abaluck, March 17th, 2019). \end{quote}

 Many years ago
\citep*{leamer_aer} in his classic paper ``Let's take the Con out of Econometrics,'' articulated this suspicion
 of models that rely on complex structures between a large number of variables
most eloquently (and curiously also in the context of studying the effect of  fertilizer on crop yields):
\begin{quote}`` The applied econometrician is like a farmer who notices that the yield is somewhat higher under trees where birds roost, and he uses this as evidence that bird droppings increase yields. However, when he presents this finding at the annual meeting of the American Ecological Association, another farmer in the audience objects that he used the same data but came up with the conclusion that moderate amounts of shade increase yields. A bright chap in the back of the room then observes that these two hypotheses are indistinguishable, given the available data. He mentions the phrase "identification problem," which, though no one knows quite what he means, is said with such authority that it is totally convincing.'' 
(\citep*{leamer_aer}, p. 31).
\end{quote}
Ultimately Leamer's concerns were part of what led to the  credibility revolution (\citep{angrist2010credibility}) with its focus on  credible identification strategies, typically in settings with a modest number (of sets) of variables.
This is why much of the training in economics PhD programs attempts to provide economists with a deep understanding of a number of the identification strategies listed earlier, regression discontinuity designs, instrumental variables, synthetic controls, unconfoundedness, and others, including the statistical methods appropriate for each of these identification strategy,  than train them to be able to infer identification in complex,  and arguably implausible, models. That is not to say that there is not an important role for structural modeling in econometrics. However, the structural models used in econometrics use economic theory more deeply than the current DAGs, exploiting monotonicity and other shape restrictions as well as  other implications of the theory that are not easily incorporated in the DAGs and the attendant \doo-calculus, despite the claims of universality of the DAGs.
 
 To be specific about the concerns about this type of DAG, let us consider adding two  causal links to Figure \ref{fig_pearl}(a), leading to Figure \ref{fig_pearl}(b). First, there is an additional direct effect of the bird population $B$ on the crop yield $Y$. Birds may eat the seeds, or parts of the plants in a way that affect the yield. There is also   a direct link from the soil fumigation $X$ on the bird population $B$: the soil fumigation may have an effect on other food sources for the birds separate from the effect on the size of the eelworm population, for example because the fumigation affects the quality of the eelworm population from the perspective of the birds. In general it is easy to come up with arguments for the presence of links: as anyone who has attended an empirical economics seminar knows, the difficult part is coming up with an  argument for the absence of such effects that convinces the audience. 
 Why is the eelworm population before the fumigation independent of the fumigation, conditional on last season's eelworm population? Why is the bird population independent of both the pre and post-fumigation eelworm population conditional on last season's eelworm population, but not independent of the end-of-season eelworm population?
 This difficulty in arguing for the absence of effects  is particularly true in social sciences where any effects that can possibly be there typically are, in comparison with physical sciences where the absence of deliberate behavior may enable the researcher to rule out particular causal links.  As 
Gelman
 puts it, ``More generally, anything that plausibly could have an effect will not have an effect that is exactly zero.'' (\citep*{gelman2011causality}, p. 961). Another question regarding the specific DAG here is why the size of the eelworm population is allowed to change repeatedly, whereas the local bird population  remains fixed.



The point of this discussion is that a major challenge in causal inference is coming up with the causal model or DAG. Establishing whether a particular model is identified, and if so, whether there are testable restrictions, in other words, the parts that a DAG is potentially helpful for, is a  secondary, and much easier, challenge.


\subsection{Some DAG Terminology}

Let me now introduce some additional terminology to facilitate the discussion. See TBOW or \citep*{pearl} for details. To make the concepts specific I will focus on the eelworm example from  Figure \ref{fig_pearl}(a). The set of {\it nodes} in this DAG is $\mmz=\{Z_0,Z_1,Z_2,Z_3,B,X,Y\}$. The {\it edges} are $Z_0\rightarrow X$, $Z_0\rightarrow Z_1$, $Z_0\rightarrow B$,
$Z_1\rightarrow Z_2$, $B\rightarrow Z_3$, 
 $X\rightarrow Y$, $X\rightarrow Z_2$, $Z_2\rightarrow Z_3$,  $Z_2\rightarrow Y$, and
  $Z_3\rightarrow Y$.
Consider a {node} in this DAG.
 For any given node, all the nodes that have arrows going directly into that node are its {\it parents}. In Figure \ref{fig_pearl}(a), for example, $X$ and  $Z_1$ are the parents of $Z_2$. 
{\it Ancestors} of a node include parents of that node, their parents, and so on.
The full set of ancestors of $Z_2$ is  $\{Z_0,Z_1,X\}$.
For any given node all the nodes that have arrows going into them directly from the  given node are its {\it children}. 
 In Figure \ref{fig_pearl}(a) $Z_2$ is the only  child of $Z_1$. 
 {\it Descendants} of a node include its children, their children, and so on. The set of descendants of $Z_1$ is $\{Z_2,Z_3,Y\}$.

 A {\it path} between two different nodes is a set of connected edges starting from the first node going to the second node, irrespective of the direction of the arrows. For example, one path going from $Z_2$ to $Z_3$ is the edge $(Z_2\rightarrow Z_3)$. Another path is $(Z_2\leftarrow X\rightarrow Y\leftarrow Z_3)$.  A {\it collider} of a path is an non-endpoint node on that path with arrows from that path going into the node, but no arrows from that path emerging from that node.  $Z_2$ is a collider on the path $(X\rightarrow Z_2\leftarrow Z_1)$. 
 Note that a node $X$ can be a collider on one path from $Z$ to $Y$ and the same node can be a non-collider on a different path from $Z$ to $Y$, so that clearly being a collider is not an intrinsic feature of a node relative to two other nodes, it also depends on the path. See Figure \ref{fig_iv}, where $X$ is a collider on the path $Z\rightarrow X\leftarrow U\rightarrow Y$ and a non-collider on the path $Z\rightarrow X\rightarrow Y$.\footnote{I thank Thomas Richardson for pointing this out to me.}
 A {\it non-collider} of a path is an non-endpoint node on a path that is not a collider. $Z_2$ is a non-collider on the path $(X\rightarrow Z_2\rightarrow Z_3)$.

Now consider types of paths. 
A {\it directed path} is a path where the arrows all go in the same direction. The path $(Z_2\rightarrow Z_3\rightarrow Y)$ is a directed path, but the path
$(Z_2\leftarrow X\rightarrow Y\leftarrow Z_3)$ is not. 
A {\it back-door path} from  node $A$ to node $B$ is a {path} from $A$ to $B$ that starts with an incoming arrow into $A$ and ends with an incoming arrow into $B$.
The path $(X\leftarrow Z_0\rightarrow Z_1\rightarrow Z_2)$ is a back-door path from $X$ to $Z_2$. A back-door path must contain at least one non-collider, although in general it may contain both colliders and non-colliders. 
A path between two nodes is {\it blocked} or {\it d-separated} by conditioning on a subset $\mmz_1$ of  the set of all nodes $\mmz$ in the DAG if and only if one of two conditions is satisfied. Either $(i)$  the path contains a noncollider that has been conditioned on, or $(ii)$ it contains a collider such that  $(a)$ that collider has not been conditioned on and $(b)$  that collider has no descendants that have been conditioned on.
In Figure \ref{fig_pearl}(a), conditioning on $\mmz_1=\{Z_2\}$ would block/d-separate the path $(X\rightarrow Z_2\rightarrow Z_3)$ because $Z_2$ is a non-collider on this path. Without conditioning on anything, $\mmz_1=\emptyset$, the path $(X\rightarrow Y\leftarrow Z_2)$ is blocked because $Y$ is a collider that is not conditioned on and that has no descendants that are conditioned on. If we condition on $\mmz_1=\{Y\}$ the path $(X\rightarrow Y\leftarrow Z_2)$ is no longer blocked because we condition on  a collider.

\subsection{The \doo-operator and the \doo-Calculus}
\label{doo}

From the joint distribution of two variables $X$ and $Y$ we can infer the conditional distribution $P(Y|X)$, and evaluate that at a particular value, say $X=x$, to get $P(Y|X=x)$. However, what we are interested in is not the distribution of the outcome we would encounter if $X$ happened to be equal to $x$, but the distribution of the outcome we would encounter if we set $X$ to a particular value. TBOW writes this using the \doo-operator as $P(Y|\doo(X=x))$ to distinguish it from the conditional distribution $P(Y|X=x)$. We can directly infer the conditional distributions of the type $P(Y|X=x)$ from the joint distribution of all the variables in the graph. Thus we take as given that we know (or can estimate consistently) all the conditional distributions $P(Y|X=x)$. The question is whether that, in combination with the assumptions embodied in the DAG, allows us to infer causal objects of the type $P(Y|\doo(X=x))$. This is what the \doo-calculus is intended to do. See 
\citet{tucci2013introduction} and TBWO for  accessible introductions, and \citep*{pearl1995causal, pearl} for more details.

How does the do-calculus relate to the DAG? Suppose we are interested in  the causal effect  of $X$ on $Y$. This corresponds to comparing $P(Y|\doo(X=x))$  for different values of $x$. To obtain
$P(Y|\doo(X))$ we  modify the graph in a specific way (we perform {\it surgery} on the graph), using the \doo-calculus. Specifically, we remove all the arrows going into $X$. 
This gives us a new causal model. For that new model the distribution $P(Y|X)$ is the same as $P(Y|\doo(X))$. So, the question is how we infer $P(Y|X)$ in the new post-surgery model from the joint distribution of all the observed variables in the old pre-surgery model. One tool is to {\it condition} on certain variables. Instead of looking at the correlation between two variables $Y$ and $X$, we may look at the conditional correlation between them where we condition on a set of additional variables. Whether the conditioning works to obtain the causal effects is one of the key questions that the DAGs are intended to answer.

The \doo-calculus has three fundamental rules. Here we use the versions in TBOW, which are slightly different from those in \citep{pearl}:
\begin{enumerate}
\item  Consider a DAG, and  $P(Y|\doo(X),Z,W)$. If, after deleting all paths into $X$, the set of variables  $Z$ blocks all the paths from $W$ to $Y$, then $P(Y|\doo(X),Z,W)=P(Y|\doo(X),Z)$.
\item If a set of variables $Z$ blocks all back-door paths from $X$ to $Y$, then $P(Y|\doo(X),Z)=P(Y|X,Z)$ (``doing'' $X$ is the same as ``seeing'' $X$).
\item  If there is no path from $X$ to $Y$ with only forward-directed arrows, then $P(Y|\doo(X))=P(Y)$.
\end{enumerate}

 Let us consider two  of the most important examples of identification strategies based on the \doo-calculus for  identifying causal effects in DAG, the {\it back-door criterion} and the {\it front-door criterion}.

\subsection{The Back-door Criterion}

The back-door criterion for identifying the causal effect of a node $X$ on a node $Y$ is based on blocking all backdoor paths through conditioning on a subset of nodes. Let us call this subset  of conditioning variables $\mmz_\bd$, where the subscript ``bd'' stands for back-door. When is it sufficient to condition on this subset? We need to check whether all back-door paths are blocked  as a result of conditioning on $\mmz_\bd$. Recall the definition of blocking or d-separating a backdoor path. It requires either conditioning on a non-collider, or the combination of not conditioning on a collider and not conditioning on all the descendants of that collider. 

Consider Figure \ref{fig_unc}. In this case conditioning on $\mmz_\bd=\{W\}$ suffices. By the second rule of the \doo-calculus, 
$P(Y|\doo(x),W)=P(Y|X=x,W)$
and so $P(Y|\doo(x))$ can be inferred from $P(Y|X=x,W)$ by integrating over the marginal distribution of $W$. In Figure \ref{fig_iv}, however, the backdoor criterion does not work. There is a backdoor path from $X$ to $Y$ that cannot be blocked, namely the path $(X\leftarrow U\rightarrow Y)$. We cannot block the path by conditioning on $U$ because $U$ is not observed.

The back-door criterion typically leads to the familiar type of statistical adjustments through matching, weighting, regression adjustments, or doubly-robust methods (see \citep{imbens2004, abadie2018econometric} for surveys). The main difference is that given the DAG it provides a criterion for  selecting the set of variables to condition on.

\subsection{The Front-door Criterion}

A second identification strategy is the front-door criterion. This strategy for identifying the effect of a variable $X$ on an outcome $Y$ does not rely on blocking all back-door paths. Instead the front-door criterion relies on the existence of intermediate variables that lie on the causal path from $X$ to $Y$.  It relies both on the effect of $X$ on these intermediate variables being identified, and on  the effect of the intermediate variables on the outcome being identified. This is an interesting strategy in the sense that it is not commonly seen in economics.

\begin{figure}
    \centering    
    \begin{subfigure}[b]{0.45\textwidth}
        \begin{tikzpicture}[
            >=stealth,
            node distance=1.2cm and 2cm            ]
            \node[observed, label=above:{smoking}] (1) {};
            \node[observed, right=of 1, label=above:{tar deposit}] (2) {};
            \node[unobserved, below=of 2, label=below:{smoking gene}] (3) {};
            \node[observed, right=of 2, label=above:{lung cancer}] (4) {};
            \draw [->, notouch] (1.east) -- (2.west);
            \draw [->, notouch] (2.east) -- (4.west);
            \draw [->, notouch] (1.east) -- (2.west);
            \draw [dashed, ->, notouch] (3.north west) -- (1.south east);
            \draw [dashed, ->, notouch] (3.north east) -- (4.south west);
        \end{tikzpicture}
        \caption{Original Pearl DAG for front-door criterion}
    \end{subfigure}
    \hfill
    \begin{subfigure}[b]{0.45\textwidth}
        \begin{tikzpicture}[
            >=stealth,
            node distance=1.2cm and 2cm
            ]
            \node[observed, label=above:{smoking}] (1) {};
            \node[observed, right=of 1, label=above:{tar deposit}] (2) {};
            \node[unobserved, below=of 2, label=below:{smoking gene}] (3) {};
            \node[observed, right=of 2, label=above:{lung cancer}] (4) {};
            \draw [->, notouch] (1.east) -- (2.west);
            \draw [->, notouch] (2.east) -- (4.west);
            \draw [->, notouch] (1.east) -- (2.west);
            \draw [dashed, ->, notouch] (3.north west) -- (1.south east);
            \draw [dashed, ->, notouch] (3.north east) -- (4.south west);
            \draw [dashed, ->, notouch] (3.north) -- (2.south);
        \end{tikzpicture}
        \caption{Freedman Concern 1: smoking gene \(\to\) tar deposits}
    \end{subfigure}
    \vspace{3em}
    \begin{subfigure}[b]{0.45\textwidth}
        \begin{tikzpicture}[
            >=stealth,
            node distance=1.2cm and 2cm
            ]
            \node[observed, label=above:{smoking}] (1) {};
            \node[observed, right=of 1, label=above:{tar deposit}] (2) {};
            \node[unobserved, below=of 2, label=below:{smoking gene}] (3) {};
            \node[observed, right=of 2, label=above:{lung cancer}] (4) {};
            \draw [->, notouch] (1.east) -- (2.west);
            \draw [->, notouch] (2.east) -- (4.west);
            \draw [->, notouch] (1.east) -- (2.west);
            \draw [dashed, ->, notouch] (3.north west) -- (1.south);
            \draw [dashed, ->, notouch] (3.north east) -- (4.south);
            \draw [->, notouch] (1.south east) to [out=-10, in=190,looseness=1.5] (4.south west);
        \end{tikzpicture}
        \caption{Freedman Concern 2: smoking \(\to\) lung cancer}
    \end{subfigure}
    \hfill
    \begin{subfigure}[b]{0.45\textwidth}
        \begin{tikzpicture}[
            >=stealth,
            node distance=1.2cm and 2cm            ]
            \node[observed, label=above:{smoking}] (1) {};
            \node[observed, right=of 1, label=above:{tar deposit}] (2) {};
            \node[unobserved, below=of 2, label=below:{smoking gene}] (3) {};
            \node[observed, right=of 2, label=above:{lung cancer}] (4) {};
            \begin{scope}[node distance= 3cm]
            \node[unobserved, right=of 3, label={[align=center]below:hazardous \\ work environment}] (5) {};
            \end{scope}        
            \draw [->, notouch] (1.east) -- (2.west);
            \draw [->, notouch] (2.east) -- (4.west);
            \draw [->, notouch] (1.east) -- (2.west);

            \draw [dashed, ->, notouch] (3.north west) -- (1.south);
            \draw [dashed, ->, notouch] (3.north east) -- (4.south);

            \draw [dashed, ->, notouch] (5.north west) -- (2.south);
            \draw [dashed, ->, notouch] (5.north east) -- (4.south);

        \end{tikzpicture}
        \caption{\label{fd_imbens} Imbens Concern: hazardous work environment}
    \end{subfigure}

  \caption{\label{fig_fd} Front-Door Criterion}
\end{figure}
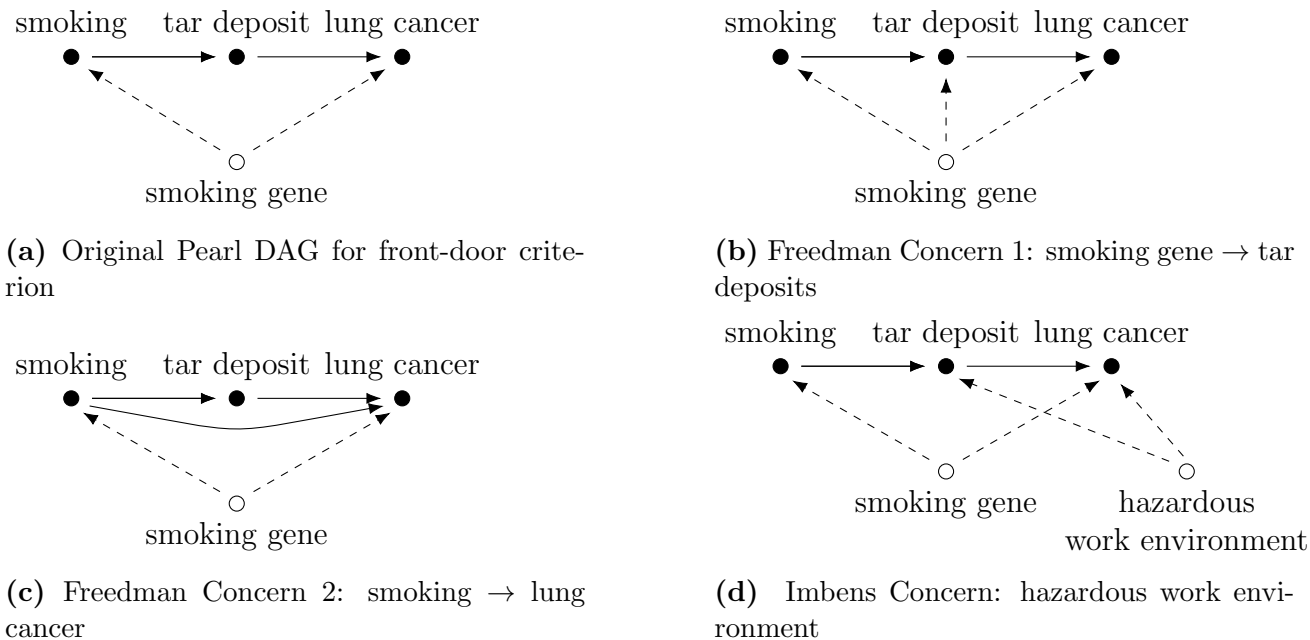

Consider the DAG in Figure \ref{fig_fd}(a) using a widely used example of the front-door criterion. We are interested in the causal effect of smoking ($X$) on lung cancer ($Y$). A smoking gene ($U$) is an unobserved confounder for this causal effect, so  $P(Y|X)\neq P(Y|\doo(X))$. However, there is another way to identify the causal effect of interest, using the additional variable tar deposits ($Z$). We can identify the causal effect of $X$ on $Z$ because there is no unobserved confounder, and $P(Z|X)= P(Z|\doo(X))$. We can identify the causal effect of $Z$ on $Y$ by adjusting for $X$, because
$P(Y|Z,X)= P(Y|\doo(Z),X)$. Putting these two together allows us to infer the causal effect of $X$ on $Y$. With discrete $Z$, the formula that relates the average causal effect of $X$ on $Y$ to the joint distribution of the realized values is
\[P(Y|\doo(x))=\sum_{x'}\sum_z P(Y|Z=z,X=x')P(x')p(Z=z|X=x).
\]
Note  the close connection to the instrumental variable DAG, given in Figure \ref{fig_iv}. Just as in  instrumental variables settings, a key assumption in the front door criterion is an exclusion restriction that the effect of one variable on another variable is mediated entirely by a third variable. In the instrumental variables case in Figure \ref{fig_iv}  it is the effect of the instrument $Z$ on the outcome $Y$ that is assumed to  be mediated by the endogenous variable  $X$. 
In the front door criterion in Figure \ref{fig_fd}(a)  it is the effect of  smoking on lung cancer that is assumed to be mediated through the tar deposit.  The connection between the instrumental variables case and the front-door criterion shows that the exclusion restriction is common to both, and thus is likely to be equally controversial in both.
The other key assumption in the instrumental variables case is that there is no unmeasured confounder for the relation between the instrument  and both the outcome and the treatment on the other hand. In many of the most convincing instrumental variables application that assumption is satisfied by design because of randomization of the instrument ({\it e.g,} the draft lottery number in  \citep*{angrist1990lifetime}, and random assignment in randomized controlled trials with noncompliance, \citep*{hirano2000assessing}). In the front door case the additional key assumption is that there are no unmeasured confounders for the intermediate variable and the outcome. Unlike the no-unmeasured-confounder assumption in the instrumental variables case this assumption cannot be guaranteed by design. As a result it will be controversial in many applications.

The front-door criterion  is an interesting case, and in some sense an important setting for the DAGs {\it versus} PO discussion. I am not aware of any applications in economics that explicitly use this identification strategy. The question arises whether this is an important method, whose omission from the canon of identification strategies  has led economists to miss interesting opportunities to identify causal effects. If so, one might argue that is should be added to the canon along side instrumental variables, regression discontinuity designs and others. TBOW clearly think so, and are very high on this identification strategy:
\begin{quote} ``[the front door criterion] allows us to control for confounders that we cannot observe ... including those we can't even name. RCTs [Randomized Controlled Trials, GWI] are considered the gold standard of causal effect estimation for exactly the same reason. Because front-door estimates do the same thing, with the additional virtue of observing people's behavior in their own natural habitat instead of a laboratory, I would not be surprised if this method eventually becomes a serious competitor to randomized controlled trials.''
(TBOW, p. 231).\end{quote}
David Cox and Nanny Wermuth on the other hand are not quite convinced, and, in a comment on \citep*{pearl1995causal} write:
\begin{quote}``Moreover, an unobserved variable $U$ affecting both $X$ and $Y$ must have no direct effect on $Z$. Situations where this could be assumed with any confidence seem likely to be exceptional.''
(\citep*{cox1995discussion}, p. 689).
\end{quote}
As the Cox-Wermuth  quote suggests, a question that does not get answered in many discussions of the front-door criterion is how credible the strategy  is in practice. The smoking and lung cancer example TBOW uses  has been used in a number of other studies as well, {\it e.g.,} \citep*{koller2009probabilistic}. TBOW mentions that the former Berkeley statistician David Freedman raised three concerns regarding  the DAG in Figure \ref{fig_fd}(a). He thought the same unobserved confounder could also affect tar deposits (as in Figure \ref{fig_fd}(b)), similar to the Cox-Wermuth concern. Smoking might also affect  cancer through other mechanisms  (as in Figure \ref{fig_fd}(c)), a concern also raised in \citep*{koller2009probabilistic}. Finally, Freedman thought the observational study would not be feasible because it is not possible to measure tar deposits in living individuals accurately. I would add to those three concerns of Freedman's a fourth, namely the  concern that there may be a second unmeasured confounder that affects tar deposits and cancer, even if it does not affect smoking, for example work environment (as in Figure \ref{fig_fd}(\subref{fd_imbens}). TBOW deflects these concerns by saying that 
\begin{quote}``I have no quarrel with Freedman's criticism in this particular example. I am not a cancer specialist, and I would always have to defer to the expert opionion on whether such a diagram represents the real-world processes accurately.'' (TBOW, p. 228)\end{quote}
Similarly \citep*{koller2009probabilistic} argue that this specific example is ``more useful as a thought experiment than a practical computational tool'' (\citep*{koller2009probabilistic}, p. 1024) and  concede that they view the substantive  assumptions as ``unlikely.''
But there's the rub. Freedman was not a cancer  specialist either, but is willing to engage with the substantive assumptions, and argues they fall short.  It is fine that a method make strong assumptions and that some of these assumptions are controversial. However, if in TBOWs favorite front-door example,  none of the authors using the example are willing to put up any defense for the substantive assumptions  that justify its use, then I agree with Cox and Wermuth that it is difficult to imagine that there is a wealth of credible examples out there waiting for the DAGs to uncover them.

The discussion of this example also appears to reflect an unwillingness to acknowledge that in many settings it is very hard to come up with convincing simple structures.  Especially in social science applications any exclusion restriction as captured by the absence of an arrow, or any absence of an unmeasured confounder without such independence guaranteed by design
is typically difficult to defend, as illustrated earlier by the Gelman quote. The recognition of this difficulty in specifying credible models  in economics  in Leamer's celebrated ``Let's take the Con out of Econometrics'' (\citet{leamer_aer}) was a big part of the motivation for the so-called credibility revolution (\citet{angrist2010credibility}) with its focus on natural experiments and clear identification.
 In a twitter discussion with Pearl,  Jason Abaluck,
 like Cox and Wermuth, questions the empirical relevance of this criterion  and  asks Judea Pearl for real-world examples where the front-door assumptions are convincing:
\begin{quote}
``Now, it might be that in addition to having a mental model of IV in their heads when they search for a "clean DAG", economists should also have a mental model of the "front-door criterion".
...
But before we get to that stage, we will need many real-world examples where the relevant assumptions seem supportable.''
(Jason Abaluck, Twitter,  25 Mar 2019).
\end{quote}

One paper that is sometimes cited in these discussions as an example of the front-door criterion is
\citep*{glynn2018front}. They analyze a job training program, with the treatment an indicator whether the individual signed up for the training program and the outcome post-program earnings. The mediator is an indicator whether the individual actually enrolled in the program. This is an interesting paper, and it illustrates the issues well, but it difficult to see it as a credible example of the front-door set up. As the authors themselves admit, 
\begin{quote}
``As we discuss in detail below, the assumptions implicit in this graph will not hold for job training programs, but this presentation  clarifies the inferential approach.'' (\citep*{glynn2018front}, p. 1042).
\end{quote}


\begin{figure}    \begin{tikzpicture}\label{napkin1}[        >=stealth,        node distance=1cm and 3cm       ]        \node[observed, label=below:{\(W\)}] (1) {};        \node[observed, right=of 1, label=below:{\(Z\)}] (2) {};        \node[observed, right=of 2, label=below:{\(X\)}] (3) {};        \node[observed, right=of 3, label=below:{\(Y\)}] (4) {};

        \node[unobserved, above=of 2, label=above right:{\(U_2\)}] (5) {};

        \begin{scope}[node distance=1.5cm and 1.5cm]
        \node[unobserved, above right=of 5, label=right:{\(U_1\)}] (6) {};
        \end{scope}
        
           \draw [dashed, ->, notouch] (6.east) to [out=145, in=90] (1.north);

        \draw [->, notouch] (1.east) -- (2.west);
        \draw [->, notouch] (2.east) -- (3.west);
        \draw [->, notouch] (3.east) -- (4.west);

        \draw [dashed, ->, notouch] (5.west) -- (1.north east);
        \draw [dashed, ->, notouch] (5.east) -- (3.north west);

        \draw [dashed, ->, notouch] (6.east) -- (4.north west);

    \end{tikzpicture}

\caption{\label{fig_new_napkin} (Based on: Figure 7.5, A new napkin problem?  TBOW, P. 240)}
\end{figure}
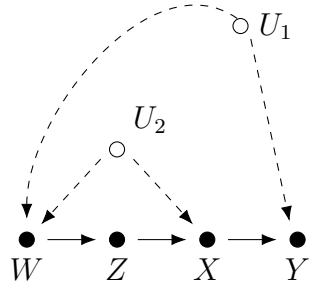

On page 240 TBOW presents one additional example, in a setting with four observed variables, linked through the path $(W\rightarrow Z\rightarrow X\rightarrow Y)$, and two unobserved confounders, $U_1$, which affects $W$ and $Y$, and $U_2$, which affects $W$ and $X$. 
See Figure \ref{fig_new_napkin}.
The question is whether the effect of $X$ on $Y$ is estimable. What is a good example where this is a natural DAG? TBOW is silent on this matter. It presents the DAG as a puzzle: can we identify the effect of $X$ on $Y$? This lack of substantive examples is exactly one of the issues that makes it hard to integrate the DAG methodology with empirical practice. 

\subsection{Mediation and Surrogates}

Analyses involving {\it mediators}, discussed in Chapter 9 in TBOW, are  common in biostatistics and epidemiology. Explicit discussions involving mediation concepts  are not as common   in economics, and probably the methods associated with them deserve more attention in this field. 
 Here I  discuss some of the basics and the related surrogacy analyses. Fundamentally mediation is about understanding causal pathways from some variable to an outcome.  See
\citep*{mackinnon2012introduction,  vanderweele2014mediation, robins1992identifiability,  pearl2014interpretation, vanderweele2015explanation, pearl2001direct, lok2016defining}.

\begin{figure}
    \centering
    \begin{subfigure}[b]{0.45\textwidth}
    \begin{tikzpicture}[
        >=stealth,
        node distance=1.5cm        ]
        \node[observed, label=above:{\(X\)}] (1) {};
        \node[observed, right=of 1, label=above:{\(S\)}] (2) {};
        \node[observed, below=of 2, label=below:{\(W\)}] (3) {};
        \node[observed, right=of 2, label=above right:{\(Y\)}] (4) {};
        \draw [->, notouch] (1.east) -- (2.west);
        \draw [->, notouch] (2.east) -- (4.west);
        \draw [->, notouch] (1.east) -- (2.west);
        \draw [->, notouch] (1.north east) to [out=55, in=125] (4.north west);
        \draw [->, notouch] (3.north west) -- (1.south east);
        \draw [->, notouch] (3.north) -- (2.south);
        \draw [->, notouch] (3.north east) -- (4.south west);
    \end{tikzpicture}    
        \caption{Mediation}
    \end{subfigure}
    \hfill
    \begin{subfigure}[b]{0.45\textwidth}
    \vspace{2em}
    \begin{tikzpicture}[
        >=stealth,
        node distance=1.5cm        ]
        \node[observed, label=above:{\(X\)}] (1) {};
        \node[observed, right=of 1, label=above:{\(S\)}] (2) {};
        \node[observed, below=of 2, label=below:{\(W\)}] (3) {};
        \node[observed, right=of 2, label=above:{\(Y\)}] (4) {};
        \draw [->, notouch] (1.east) -- (2.west);
        \draw [->, notouch] (2.east) -- (4.west);
        \draw [->, notouch] (1.east) -- (2.west);
        \draw [->, notouch] (3.north west) -- (1.south east);
        \draw [->, notouch] (3.north) -- (2.south);
        \draw [->, notouch] (3.north east) -- (4.south west);
    \end{tikzpicture}   
        \caption{Surrogates}
    \end{subfigure}
    \vspace{3em}
    \begin{subfigure}[b]{0.45\textwidth}
    \vspace{2em}
    \begin{tikzpicture}[
        >=stealth,
        node distance=1.5cm        ]
        \node[observed, label=above:{\(X\)}] (1) {};
        \node[observed, right=of 1, label=above right:{\(S\)}] (2) {};
        \node[observed, below=of 2, label=below:{\(W\)}] (3) {};
        \node[observed, right=of 2, label=above:{\(Y\)}] (4) {};        
        \node[unobserved, above right=of 1, label=above:{\(U\)}] (5) {};
        \draw [->, notouch] (1.east) -- (2.west);
        \draw [->, notouch] (2.east) -- (4.west);
        \draw [->, notouch] (1.east) -- (2.west);
        \draw [->, notouch] (3.north west) -- (1.south east);
        \draw [->, notouch] (3.north) -- (2.south);
        \draw [->, notouch] (3.north east) -- (4.south west);
        \draw [dashed, ->, notouch] (5.south west) -- (1.north east);
        \draw [dashed, ->, notouch] (5.south east) -- (2.north west);
    \end{tikzpicture}    
        \caption{Invalid Surrogates}
    \end{subfigure}
    \hfill
    \begin{subfigure}[b]{0.45\textwidth}\vspace{2em}
    \begin{tikzpicture}[
        >=stealth,
        node distance=1.5cm        ]
        \node[observed, label=above:{\(X\)}] (1) {};
        \node[observed, right=of 1, label=above:{\(S\)}] (2) {};
        \node[observed, below=of 2, label=below:{\(W\)}] (3) {};
        \node[observed, right=of 2, label=above right:{\(Y\)}] (4) {};        
        \node[unobserved, above right=of 2, label=above:{\(U\)}] (5) {};
        \draw [->, notouch] (1.east) -- (2.west);
        \draw [->, notouch] (2.east) -- (4.west);
        \draw [->, notouch] (1.east) -- (2.west);
        \draw [->, notouch] (3.north west) -- (1.south east);
        \draw [->, notouch] (3.north) -- (2.south);
        \draw [->, notouch] (3.north east) -- (4.south west);
                \draw [dashed, ->, notouch] (5.south west) -- (2.north east);
        \draw [dashed, ->, notouch] (5.south east) -- (4.north west);
    \end{tikzpicture}
        \caption{Invalid Surrogates}
    \end{subfigure}
  \caption{\label{mediation} Mediation and Surrogacy}
\end{figure}
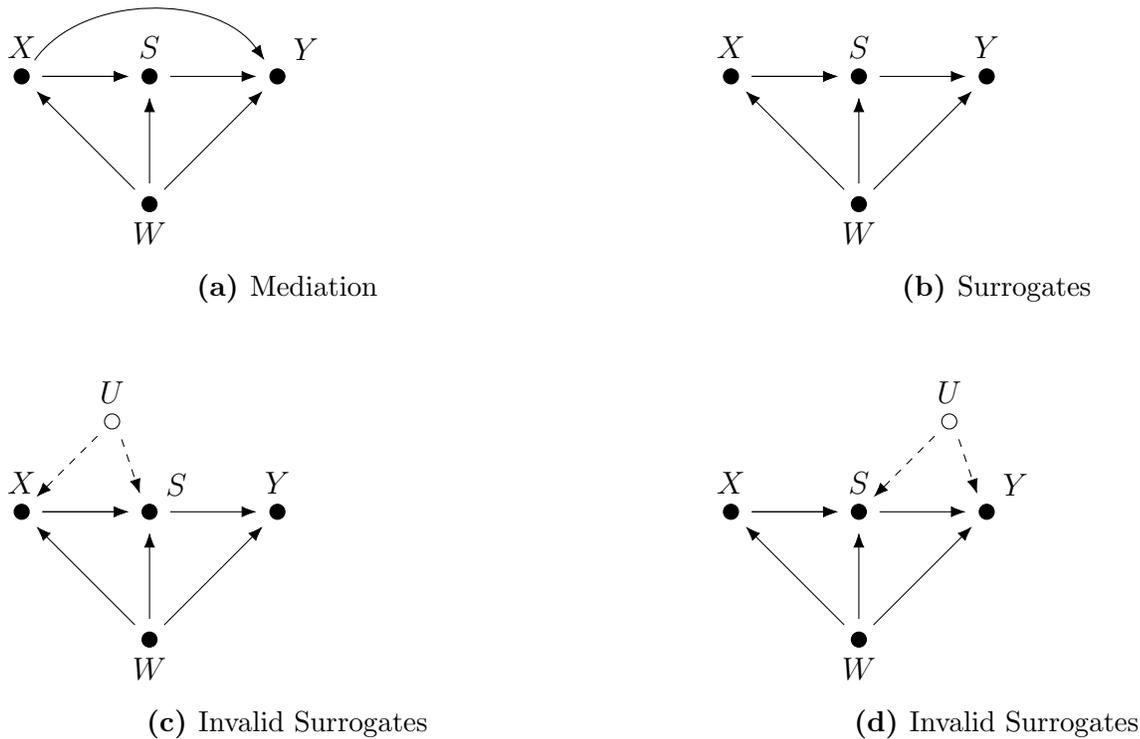

Let us consider a specific example from \citep*{vanderweele2015explanation}. There is a well-established association between some genetic variances on chromosome 15q25.1 and lung cancer. Part of this link may be through smoking. So, in this case we are interested in understanding the role of a potential mediator, smoking $(S)$ in the causal link between the genetic variation ($X)$ on lung cancer ($Y)$. There may be an observed confounder $W$ that affects the variable of interest, the genetic variant, the mediator, smoking, and the outcome, lung cancer.
Figure \ref{mediation}(a) presents the basic mediation case. There is a direct causal link from the basic treatment to the outcome, as well as a link from the basic treatment to the mediator and from the mediator to the outcome. In this case, with no unobserved confounders complicating matters, we can infer all the causal effects of interest, and we can separate out the direct effect of the genetic variation and the indirect effect that is mediated through smoking. 
Let us be more specific. There are three effects we are interested in. First, the total effect of the genetic variation on lung cancer. We can identify that total effect given the DAG because there are no back-door paths. Second, the indirect effect of the genetic variation on lung cancer. This consists of two components, the effect of genetic variation on smoking, and the effect of smoking on lung cancer. Given the DAG we can estimate the first of these two because there is no back-door path. We can also estimate the second, the effect of smoking on lung cancer by controlling for the genetic variation. Third, the direct effect, which we can infer by subtracting the indirect effect from the total effect. See  \citep*{vanderweele2015explanation} for details, an exposition of the precise definition of the estimands using the PO approach.

The value of the mediation analysis is that it sheds light on the causal pathways. The challenge is that it requires us to identify a number of constituent causal effects. As such it requires more assumptions than those required for estimating the total effect alone. However, if those assumptions are satisfied, it delivers more than simply the total effect.

A closely related set up is that of surrogate variables. The basic DAG is shown in
Figure \ref{mediation}(b). Compared to the mediation set up a key additional assumption is that there is no direct effect in this case. Under the same assumptions as required for the mediation case this would lead to testable restrictions, but the typical  use case is a different one. As in 
\citep*{athey2016estimating}, a prominent use case is that with two samples  (see also
\citep*{gupta2019top} discussing the use of two-sample surrogacy methods in the context of experimentation in tech companies). The first sample comes from a randomized experiment where both the basic treatment and the surrogates/mediators are observed (possibly plus some pre-treatment variables), but not the outcome of interest. The second sample is from an observational study where only the surrogates and the primary outcome are observed (possibly plus the pre-treatment variables), but not the basic treatment. The goal is to estimate the causal effect of the treatment on the outcome, without ever seeing data on the the treatment and the outcome for the same units. In practice the setting often involves multiple surrogates. For example, in an online experimental setting one may observe many intermediate outcomes in a short term experiment, without observing the long term outcome of interest.

In the surrogate case, as in the mediation case, the DAG can clarify the content of the assumptions. In particular it rules out a direct effect of the treatment on the outcome (as in Figure \ref{mediation}(a)). It also rules out unobserved confounders that affect both the treatment and the surrogates (as in 
Figure \ref{mediation}(c)). Finally, it rules out by assumption unobserved confounders that affect both the surrogates and the primary outcome (as in 
Figure \ref{mediation}(d)).
The DAGs have less to offer in terms of estimation strategies and formal statistical assumptions in this setting. Here the PO framework offers clear estimation strategies and methods for inference. See 
\citep*{athey2016estimating} for details. Note also the close connection between the surrogate DAG and the instrumental variables DAG.

\subsection{``Elements of Causal Inference,'' (\citep*{peters2017elements}) and Other New Developments}

\citep*{peters2017elements} is a fascinating book discussing recent new directions for causal inference followed in the computer science literature, often using graphical models. Many of the problems studied in this book are quite different from those studied traditionally in the economics literature. For example, there is much interest in this literature in assessing the direction of causality, whether $X$ causes $Y$ or $Y$ causes $X$. 
This question received considerable attention in the econometric literature in the context of time series data, leading to the concept of Granger-Sims causality (\citep*{granger1969investigating, sims1972money, chamberlain1982general}). In contrast, the CS literature is focused on a cross-section setting, where we have observations on exchangeable pairs $(X_i,Y_i)$. 
Consider for example two linear models:
\[ Y_i=\alpha_0+\alpha_1 X_i+\varepsilon_i,
\hskip1cm {\rm and}\ \ 
 X_i=\beta_0+\beta_1 Y_i+\eta_i.\]
Can we tell which of these models is the right one? Obviously without additional assumptions we cannot, but if we are willing to put additional structure on the model we may be able to make progress. For example, \citep*{peters2017elements} considers the assumption that the unobserved term ($\varepsilon_i$
or $\eta_i$) is independent of the right-hand side variable. This is still not sufficient for choosing between the models if the distributions are Gaussian, but outside of that, we can now tell the models apart.
Identifying models based on this type of functional form and distributional assumption is not common in economics. The basic question is also an unusual one in economics settings. In many economic settings we know the cause and the outcome, that is, we know the direction of the causality, the questions are primarily about the magnitude of the causal effects, and the possible presence of unmeasured confounders. For example we are interested in the effect of education on earnings, not the effect of earnings on education because we know which comes first. In cases where we are unsure about the direction of the causality there is typically more structure on the problem as well, in terms of mechanisms involving expectations and feedback loops so that the researcher is not simply trying to infer from a data set on $(X_i,Y_i)$ the direction of the causality.

This example shows how the questions studied in this literature are different from those in economics. This is particularly true for the questions on causal discovery 
(\citep*{uhler2013geometry, glymour2014discovering, hoyer2009nonlinear, lopez2017discovering, mooij2016distinguishing, dubois2017causal}), where the goal is to find causal structure in data, without starting with a fully specified model. The fact that the questions in this literature are currently quite different from those studied in economics  does not take away from the fact that ultimately the results in this literature may be very relevant for research in social sciences. The aim in the causal discovery literature is to infer from complex data directly the causal structure governing these data. If succesful, this would be very relevant for social science questions, but it is currently not there yet.

More closely related to the econometric literature is the recent work on invariance and causality (\cite{peters2016causal, buhlmann2018invariance}).

\section{Potential Outcomes and the Rubin Causal Model}
\label{section:po}

What I refer to here as the PO framework is what  \citep*{holland1986statistics} calls the Rubin Causal Model. 
 \citep*{rubin1974estimating} is a very clear and non-technical introduction, and
\citep*{imbens2015causal}
is a textbook treatment. 
It has many antecedents in the econometrics literature, as early  as the 1930s and 1940s and it is currently widely used in the empirical economics literature. Here I give a brief overview.
The starting point is  a population of units. There are then three components of the PO approach. First, there is a treatment/cause that can take on different values for each unit. Each unit  in the population is characterized by a set of potential outcomes $Y(x)$, one for each level of the treatment. In the simplest case with a binary treatment there are two potential outcomes, $Y(0)$ and $Y(1)$, but in other cases there can be more. Only one of these potential outcomes can be observed, namely the one corresponding to the treatment received:
\[ Y^\obs=Y(X)=\sum_{x} Y(x){\bf 1}_{X=x}.\]
 The others are {\it ex post} counterfactuals. 
The causal effects correspond to comparisons of the potential outcomes, of which at most one can be observed, with all the others missing. Paul Holland refers to this as the ``fundamental problem of causal inference,'' (\citep*{holland1986statistics}, p. 59).
This leads to the second component, the presence of multiple units so that we can see units receiving each of the variour treatments.
 The third key component is the assignment mechanism that determines which units receive which treatments.
Much of the literature has concentrated on the case with just a single binary treatment, with the focus on estimating the average treatment effect of this binary treatment for the entire population or some subpopulation
\[ \tau=\mathbb{E}[Y_i(1)-Y_i(0)].\]
In the \doo-calculus, this would be written as $\tau=\mathbb{E}[Y(\doo(1)fs)-Y(\doo(0))].$
At this stage the difference between the PO and DAG approach is relatively minor.
The disctinction between $P(Y|X=x)$ and $P(Y|\doo(x))$ in the PO framework corresponds to the difference between $P(Y^\obs|X=x)$ and $P(Y(x))$.
There is a smaller literature on the case with discrete or continuous treatments ({\it e.g.,} \citep*{imbens2000}).

\subsection{Potential Outcomes and Econometrics}

For each unit, and for each value of the treatment, there is a potential outcome that could be observed if that unit was exposed to that level of the treatment. 
We cannot see the set of potential outcomes for a particular unit because each unit can be exposed to at most one level of the treatment, and only the potential outcome corresponding to that level of the treatment can ever be observed. With a single unit and a binary treatment, the two potential outcomes could be labelled $Y(0)$ and $Y(1)$, with the causal effect being a comparison of the two, say, the difference $Y(1)-Y(0)$.
This is a simple, but powerful notion. It has resonated with the econometrics and empirical economics community partly because it directly connects with the way economists think about, say, demand and supply functions. The notion of potential outcomes is very clearly present in the work of  Wright, Tinbergen and Haavelmo in the 1930s and 1940s (\citep*{wright1928tariff, tinbergen1930determination,  haavelmo1943statistical}).
For example, Tinbergen carefully distinguishes in his notation between the price as an argument in the supply and demand function, and the realized equilibrium price:
\begin{quote}
``Let $\pi$ be any imaginable price; and call total demand at this price $n(\pi)$, and total supply $a(\pi)$. Then the actual price $p$ is determined by the equation 
\[ a(p)=n(p),\]
so that the actual quantity demanded, or supplied, obeys the condition $u=a(p)=n(p)$, where $u$ is this actual quantity.`' (\citep*{tinbergen1930determination}, translated in \citep*{hendry1997foundations}, p.  233)
\end{quote}
Note also that the potential outcomes, here the supply and demand function, are taken as primitives, in line with much of the economic literature. 
Similar to Tinbergen,  Haavelmo writes:
\begin{quote}``Assume that if the group of all consumers in the society were repeatedly furnished with the total income, or purchasing power, $r$ per year, they would, on the average, or `normally', spend a total amount $\bar u$ for consumption per year, equal to
\[\bar u=\alpha r+\beta,
\] where $\alpha$ and $\beta$ are constants. The amount, $u$ {\it actually} spent each year might be different from $\bar u$.`' (italics in original, \citep*{haavelmo1943statistical}, p. 3)
\end{quote}
Some of the clarity of the potential outcomes that is present in \citep*{tinbergen1930determination} and \citep*{haavelmo1943statistical}  got lost in some of the subsequent econometrics. The Cowles Foundation research which led to the general simultaneous equations set up, modeled only the observed outcomes and dropped the notation for the potential outcomes. The resurgence of the use of potential outcomes in the econometric program evaluation literature starting with \citep*{heckman1990varieties} and \citep*{manski_bounds} caught on  precisely because of the precursors in the earlier econometric literature.

\subsection{The Assignment Mechanism}

The second key component of the PO approach is the assignment mechanism. Given the multiple potential outcomes for a unit, there is only one of these that can be observed, namely the realized outcome corresponding to the treatment that was received. Critical is how the treatment was chosen, that is the assignment mechanism, as a function of the pretreatment variables and the potential outcomes.. In the simplest case, that of a completely randomized experiment, it is known to the researcher how the treatment was determined. The assignment mechanism does not depend on the potential outcomes, and it has a known distribution. For this case we understand the critical analyses well. This can be relaxed by assuming only unconfoundedness, where the assignment mechanism is free of dependence on the potential outcomes, but can depend in arbitrary and unknown ways on the pretreatment variables. Again there is a huge literature with many well-understood methods. See \citep*{imbens2004, imbens2015causal, rubin2006matched, abadie2018econometric} for recent reviews. The most complicated case is where selection is partly on unobservables, and this case has received the most attention in the econometrics literature because the starting point there is that agents make deliberate choices, rather than receiving their treatments by chance (\citet{imbens2014}).

\subsection{Multiple Units and Interference}

In many analyses researchers assume that there is no interference between units, part of what Rubin calls SUTVA (stable unit treatment value assumption).  This greatly simplifies analyses, and makes it conceptually easier to separate the questions of identification and estimation. However, the assumption that there is no interference is in many cases implausible. There is also a large and growing literature analyzing settings  where this assumption is explicitly relaxed. Within the PO framework this is conceptually straightforward. Key papers include
\citep*{manski1993,  manski2013identification, hudgens, athey2018exact, aronow2012, aronowsamii, basse2019randomization, sobel2006randomized}. In many of the settings considered in this literature it is not so clear what the joint distribution is that can be estimated precisely in large samples. As a result the separation between identification of the distribution of observed variables and the identification of causal effects that underlies many of the DAG analyses is no longer so obvious.
Recent work using DAGs with interference includes \citep*{ogburn2014causal}.

\subsection{Randomized Experiments and Experimental Design}

In the PO literature there is a very special place for randomized experiments.
 Consider the  simplest such setting, where $N_t$ units out of the population are randomly selected to receive the treatment and the remaining $N_c=N-N_t $ are assigned to the control group, and where the no-interference assumption holds. The implication of the { experimental design} is that the treatment is independent of the potential outcomes, or
\[ W_i\ \indep\ \Bigl(Y_i(0),Y_i(1)\Bigr).\]
This validates simple estimation strategies. For example, it
implies that
\[ \hat\tau=\overline{Y}_t-\overline{Y_c},\hskip1cm {\rm where}\ 
\overline{Y}_t=\frac{1}{N_t}\sum_{i:W_i=1} Y_i,\hskip0.5cm {\rm and}\ \ 
\overline{Y}_c=\frac{1}{N_c}\sum_{i:W_i=0} Y_i,
\]
is unbiased for the average treatment effect, and that 
\[ \frac{1}{N_t(N_t-1)}\sum_{i:W_i=1}\left(Y_i-\overline{Y}_t\right)^2+
\frac{1}{N_c(N_c-1)}\sum_{i:W_i=0}\left(Y_i-\overline{Y}_c\right)^2,\]
is a conservative estimator for the variance, originally proposed by \citep{neyman1923}.


The primacy of randomized experiments 
has long  resonated with economists, despite the limited ability to actually do randomization.
Since the late 1990s the development economics literature has embraced the strength of experiments
(\citep*{banerjee2008experimental}), leading to a huge empirical literature that has had a major influence on policy. \citep*{angrist2010credibility} quote \citep*{haavelmo1944} as arguing that we should at least have such an experiment in mind, even if we are not actually going to (be able to) conduct one:
\begin{quote}
``Over 65 years ago, Haavelmo submitted the following complaint to the readers of Econometrica (1944, p. 14): “A design of experiments (a prescription of what the physicists call a ‘crucial experiment’) is an essential appendix to any quantitative theory. And we usually have some such experiment in mind when we construct the theories, although--unfortunately--most economists do not describe their design of experiments explicitly.'' (\citep*{angrist2010credibility}, p. 16)
\end{quote}
Similarly,
the statistics literature is of course full of claims that randomized experiment are the most credible setting for making causal claims.
\citep*{freedman2006statistical} for example is unambiguous about the primacy of RCTs:
\begin{quote} ``Experiments offer more reliable evidence on causation than observational studies" (\citep*{freedman2006statistical}, abstract).\end{quote}

When going beyond randomized experiments,
researchers in the PO framework often analyze  observational studies by viewing them as {emulating}  particular randomized experiments, and analyzing them as if there was approximately a randomized experiment.
The {\it natural experiment} and credibility revolution literatures (\citep*{angrist2010credibility}), and much of the subsfsequent empirical literature, are focused on finding settings where assignment is as good as random at least for a subpopulation, using the various identification strategies including matching, regression discontinuity designs, synthetic control methods, and instrumental variables.
In contrast, 
the graphical literature is largely silent about experiments, and does not see them as special.
In fact, Pearl proudly proclaims himself a skeptic of any superiority of RCTs.
When Angus Deaton and Nancy Cartwright write ``We argue that any special status for RCTs is unwarranted." (\citep*{deaton2018understanding}, page 2),
Pearl comments that 
\begin{quote}``As a veteran skeptic of the supremacy of the RCT, I welcome D\&C's challenge wholeheartedly.''
(\citep*{pearl2018challenging})
\end{quote}

If anything the importance of RCTs has increased in recent years, both in academic circles, as well as outside in the tech companies:
``Together these organizations [Airbnb, Amazon,   Booking.com,   Facebook,   Google,   LinkedIn,   Lyft, Microsoft,    Netflix,    Twitter,    Uber,    Yandex,    and    Stanford University, GWI]  tested more than one hundred  thousand  experiment  treatments  last  year''  (\citep*{gupta2019top}, p.20). This has spurred a new literature on the design of experiments in complex environments. 
There are now many computer scientists, economists and statisticians working on complex experimental designs that take account of interference (\citep*{hudgens, athey2018exact, aronow2012, aronowsamii, basse2019randomization}, and that use multi-armed bandit methods and other adaptive designs, 
\citep*{scott2010modern, dimakopoulou2017estimation, dimakopoulou2018balanced}.
Even in the traditional setting of RCTs for medical treatments these issues have led to new and innovative designs, {\it e.g.,}
\citep*{isakov2019fda, das2017re}.

\subsection{Unconfoundedness}

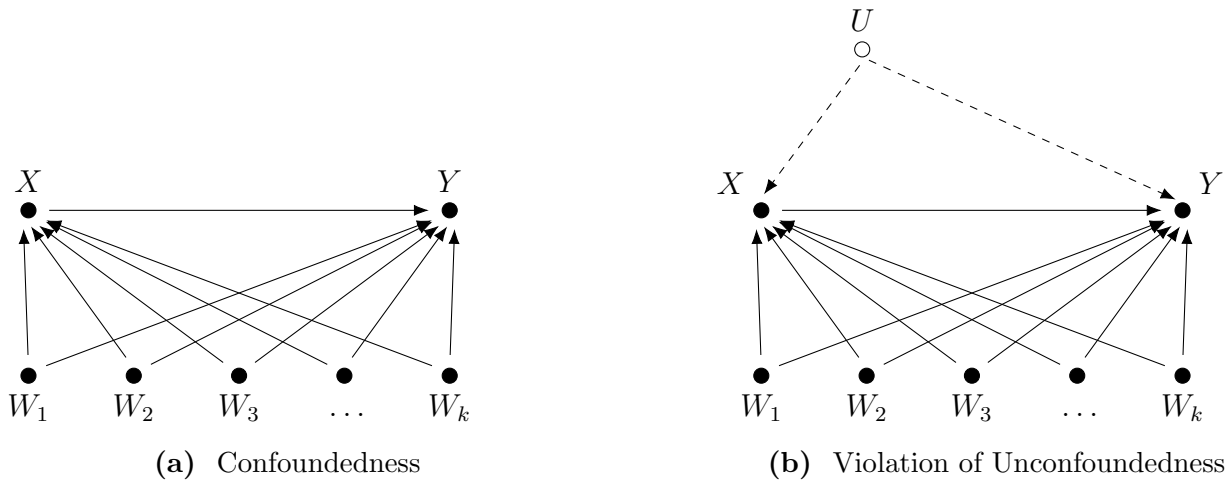
\begin{figure}
    \centering    
    \begin{subfigure}[b]{0.45\textwidth}    
    \begin{tikzpicture}[
        >=stealth,
        node distance=2cm and 1.2cm
        ]
        \node[observed, label=below:{\(W_1\)}] (1) {};
        \node[observed, right=of 1, label=below:{\(W_2\)}] (2) {};
        \node[observed, right=of 2, label=below:{\(W_3\)}] (3) {};
        \node[observed, right=of 3, label=below:{\(\vphantom{X}\dots\)}] (4) {};
        \node[observed, right=of 4, label=below:{\(W_k\)}] (5) {};
        \node[observed, above=of 1, label=above:{\(X\)}] (6) {};
        \node[observed, above=of 5, label=above:{\(Y\)}] (7) {};
        \draw [->, notouch] (6.east) -- (7.west); 
        \draw [->, notouch] (1.north) -- (6.south west);
        \draw [->, notouch] (1.north east) -- (7.south west);
        \draw [->, notouch] (2.north) -- (6.south west);
        \draw [->, notouch] (2.north east) -- (7.south west);
        \draw [->, notouch] (3.north west) -- (6.south);
        \draw [->, notouch] (3.north east) -- (7.south);
        \draw [->, notouch] (4.north west) -- (6.south east);
        \draw [->, notouch] (4.north) -- (7.south east);
        \draw [->, notouch] (5.north west) -- (6.south east);
        \draw [->, notouch] (5.north) -- (7.south east);
    \end{tikzpicture}
    
        \caption{\label{unconf} Confoundedness}
    \end{subfigure}
    \hfill
    \begin{subfigure}[b]{0.45\textwidth}

    \vspace{2em}
    \begin{tikzpicture}[
        >=stealth,
        node distance=2cm and 1.2cm        ]
        \node[observed, label=below:{\(W_1\)}] (1) {};
        \node[observed, right=of 1, label=below:{\(W_2\)}] (2) {};
        \node[observed, right=of 2, label=below:{\(W_3\)}] (3) {};
        \node[observed, right=of 3, label=below:{\(\vphantom{X}\dots\)}] (4) {};
        \node[observed, right=of 4, label=below:{\(W_k\)}] (5) {};

        \node[observed, above=of 1, label=above left:{\(X\)}] (6) {};
        \node[observed, above=of 5, label=above right:{\(Y\)}] (7) {};

        \node[unobserved, above right=of 6, label=above:{\(U\)}] (8) {};

        \draw [->, notouch] (6.east) -- (7.west); 

        \draw [->, notouch] (1.north) -- (6.south west);
        \draw [->, notouch] (1.north east) -- (7.south west);
        \draw [->, notouch] (2.north) -- (6.south west);
        \draw [->, notouch] (2.north east) -- (7.south west);
        \draw [->, notouch] (3.north west) -- (6.south);
        \draw [->, notouch] (3.north east) -- (7.south);
        \draw [->, notouch] (4.north west) -- (6.south east);
        \draw [->, notouch] (4.north) -- (7.south east);
        \draw [->, notouch] (5.north west) -- (6.south east);
        \draw [->, notouch] (5.north) -- (7.south east);
        
        \draw [dashed, ->, notouch] (8.south east) -- (6.north west);
        \draw [dashed, ->, notouch] (8.south west) -- (7.north east);

    \end{tikzpicture}   
        \caption{\label{unconf_viol} Violation of Unconfoundedness}
    \end{subfigure}

  \caption{\label{fig_unc_many} Unconfoundedness with Multiple Observed Confounders}
\end{figure}

One of the most common settings for estimating treatment effects is that under unconfoundedness or ignorable treatment assignment. The key assumption is that given a set of pre-treatment variables it is assumed that assignment to treatment is independent of the potential outcomes.
\[ X_i\ \indep \ \Bigl(Y_i(0),Y_i(1)\Bigr)\ \Bigl|\ W_i.\]
A DAG representing this set up is given in Figure \ref{fig_unc_many}(\subref{unconf}).

The DAG here does not include additional links between the pretreatment variable, but the estimation strategies justified by the DAG as given can allow for arbitrary links between them.
It does rule out the presence of an unobserved confounder, {\it e.g.,} the variable $U$ in Figure \ref{fig_unc_many}(\subref{unconf_viol}).
Because it is such a canonical setting, the subject of a huge theoretical literature as well as the basis for a vast empirical literature, it would appear that the form of the assumptions, graphical or algebraic, would no longer be a concern in practice because researchers understand what is the issue here, and can move easily between the DAG and the PO form. However, TBOW questions whether researchers understand these assumptions:
\begin{quote}``Unfortunately, I have yet to find a single person who can explain what ignorability means in a language spoken by those who need to make this assumption or assess its plausibility in a given problem.''
(TBOW, p. 281).
\end{quote}
The unconfoundedness assumption implies that the average treatment effect
$\tau=\mathbb{E}[Y_i(1)-Y_i(0)]$
 can be written  in terms of the joint distribution of the observed variables $(Y_i,X_i,W_i)$ as
\[ \tau=\mathbb{E}[Y_i(1)-Y_i(0)]=\mathbb{E}\Bigl[\mathbb{E}[Y_i|X_i=1,W_i]-\mathbb{E}[Y_i|X_i=0,W_i]\Bigr].\]
As arguably the most common setting in empirical work for estimating treatment effects, there is a huge theoretical literature on specific methods for estimating the average effect of the treatment on the outcome in this case that has closely interacted with the similarly vast empirical literature.
All methods involve adjusting in some fashion for the difference in pretreatment values $W_i$ between treated and control units. 
The methods differ in terms of the components of the joint  distribution of the observed variables they focus on.
 A key paper is \citep*{rosenbaum1983central}, and more generally the papers in \citep*{rubin2006matched}.
 A key result is that irrespective of the number of pre-treatment variables, it is sufficient to adjust for differences between treated and control units in just a scalar function of the pre-treatment variables, the propensity score, defined as $e(w)={\rm pr}(X_i=1|W_i=w)$.
Much of the statistical and econometric literatures have focused on 
 effective ways of implementing these ideas in statistics and econometrics. 
 Common in practice is still the simple least squares estimator, but much of the methodological literature has developed more robust and efficient methods.
 Some focus on estimating $\mu(x,w)=\mathbb{E}[Y_i|X_i=x,W_i=w]$ followed by averaging the difference $\hat\mu(1,W_i)-\hat\mu(0,W_i)$ over the sample distribution of the pre-treatment variables $W_i$. Another strand of the literature has focused on inverse propensity score weighting, with the weight for unit $i$ proportional  to $(e(W_i)^{X_i}(1-e(W_i))^{1-X_i})^{-1},$ where $e(w)={\rm pr}(X_i=1|W_i=x)$ is the propensity score (\citep*{rosenbaum1983central}). The current state of the literature is that the most effective and robust methods use estimates of the conditional outcome means as well as estimates of the propensity score, in what is referred to as {\it doubly robust methods}.
 See \citep*{robins1994estimation, imbens2009recent, abadie2018econometric} for  surveys.
 
Recently this literature has focused on the case with a relatively large number of pre-treatment variables (\citet{chernozhukov2017double, athey2018approximate, van2011targeted, shi2019adapting}). 
Overlap issues in covariate distributions, absent in all the the DAG discussions, become prominent among practical problems (\citep*{crump2009dealing, d2017overlap}).
 In this setting there has been much interaction with the machine learning literature, focusing on regularization methods that are effective given the particular causal estimand, rather than effective for prediction purposes.
In this setting simply assuming that one knows or can consistently estimate the joint distribution of all variables in the model, which is the basis of the discussion in TBOW and \citep*{pearl}, is not helpful, and the corresponding ``statistical  vs. causal dichotomy'' (\citep*{pearl}, p. 348) becomes blurred.

\subsection{Sensitivity Analyses}

The PO framework has been a natural setting for considering violations of the assumptions required for point identification. For example, starting with an unconfoundedness  setting, one may wish to consider the presence of an unobserved confounder that has some limited effect on the treatment assignment as well as on the outcome, and assess the sensitivity of the results to the presence of such confounders (\citep*{rosenbaum1983assessing, imbens2003}).
More recent work has taken these sensitivity analyses further, {\it e.g.,}
\citep*{andrews2017measuring, andrews2019simple}.

\section{Graphical Models, Potential Outcomes and Empirical Practice in Economics}
\label{section:practice}

Since its inception in the late 1980s and early 1990s the DAG approach to causality has generated much interest in among others, computer science, epidemiology, and parts of social science ({\it e.g.,} in sociology,
\citep*{morgan2014counterfactuals}). 
In epidemiology in particular it has connected well with related work on structural equations models  (\citep*{greenland1999causal, robins1997causal}).
 It has not, however, made major inroads into the theoretical econometric or empirical economic literatures, and is absent from most econometric textbooks ({\it e.g.,} \citep{wooldridge2001} and MHE). Although causal inference broadly defined has a long and prominent tradition in econometrics, and since the 1990s there has been a sharp increase in work explicitly focused on this area, this work  has relied primarily on the PO framework.
There are exceptions to this, and there are now some discussions of DAGs in the econometric literature ({\it e.g.,} the  discussion in
\citep*{mixtape}), and some links in the statistics literature (\citep{richardson2011transparent, malinsky2019potential}) but these remain the exception. Much of the empirical work  in economics
is closer to the PO  approach.
Pearl has discussed this, but dismissed the possibility that there are substantive reasons for this:
\begin{quote} ``Or, are problems in economics different from those in epidemiology? I have examined the structure of typical problems in the two fields, the number of variables involved, the types of data available, and the nature of the research questions. The problems are strikingly similar.'' \citep{pearl2014blog}.
\end{quote}

In this part of this essay I will explore this issue further.
 I think the case for the DAGs is the strongest in terms of exposition. The DAGs are often clear and accessible ways to expressing visually some, though not necessarily all, of the key assumptions. It may provide the reader, even more than the researcher, with an easy way to interpret and assess the overall model. I am less convinced that the formal identification results are easier to derive in the DAG framework for the type of problems commonly studied in economics. I see three reasons for that. First, the DAGs  have difficulty coding shape restrictions such as monotonicity and  identification results for subpopulations. Second, the advantages of the formal methods for deriving identification results with DAGs are most pronounced in complex  models with many variables that are not particularly popular in empirical economics.
Third, the PO approach has connected well with estimation and inference issues. Although the PO approach has largely focused on the problem of estimating average effects of binary treatments, it has been able to make much progress there, not just in terms of identification, but also on problems regarding study design, estimation and inference.
On this issue that the PO or Rubin-Causal-Model framework is much more closely tied to statistics and practical issues in inference,  Steve Powell writes in his review of TBOW that
\begin{quote} ``Donald Rubin offers an alternative framework  [the PO framework, GWI], also not shy of dealing with causality, which is much more fully developed for the needs and concerns of social scientists in general and evaluators in particular.'' (\citep*{powell_pearl}, p. 53)
\end{quote}

I will focus in this section on six specific issues.
First, I will discuss the role of randomized experiments and the importance of manipulability. Second, I will discuss instrumental variables, the importance of shape restrictions in economics. Third, I will discuss the role of simultaneity in economics. Fourth, I discuss unconfoundedness, M-bias, and the choice of the set of conditioning variables.
Fifth, I will discuss the difficulty with counterfactuals. Finally, I will discuss identification strategies in the context of the returns to education.

\subsection{Non-manipulable Causes or Attributes, Hypothetical Experiments, and the Role of the Assignment Mechanism}

The PO framework starts by defining the potential outcomes with reference to a {\it manipulation} ({\it e.g.,} \citep*{imbens2015causal}, p. 4). 
In doing so it makes a distinction between {\it attributes} or pre-treatment variables which are fixed for the units in the population, and {\it causes}, which are potentially manipulable. This is  related to the connection between causal statements and randomized experiments. The causes are conceptually tied to,  at the very least  hypothetical, experiments.
This may appear to be a disadvantage as it leads to difficulties  in the PO framework when making causal statements about such attributes as race or gender. See the discussion in \citep{holland1986statistics, doi:10.1080/01621459.1986.10478358}. In the modern causal literature in economics researchers have often acknowledged this difficulty and  focused on causal effects tied to manipulable aspects of attributes. A seminal example is \citep*{bertrand2004emily} who study the well-defined causal  effect of manipulation of the perception of race by changing names from Caucasian sounding to African-American sounding ones, rather than the less well-defined causal effect of race itself. 
This view concerning the importance of manipulability is shared in part of the statistics literature, where David Cox writes,
\begin{quote}``In this discussion, only those variables which in the context in question can conceptually be manipulated are eligible to represent causes, {\it i.e.,} it must make sense, always in the context in question, that for any individual the causal variable might have been different from the value actually taken. Thus in most situations gender is not a causal variable but rather an intrinsic property of the individual.'' 
(\citep*{cox1992causality}, p. 296).
\end{quote}
The advantage of being explicit about which variables are causal and which are not is that on only needs to consider the determinants (parents in the DAG terminology) of the causes, and not those of the attributes. 

In contrast,
Pearl sees no difficulty in defining causal effects for non-manipulable attributes in a DAG using the \doo-calculus:
\begin{quote}
``Another misguided doctrine denies causal character to nonmanipulable variables.'' (\citep*{pearl2015trygve} p. 172).
\end{quote}
This is not simply a matter of impicitly assuming that the effect of such a variable is invariant to the method in which is it manipulated, and that therefore the articulation of the manipulation is immaterial.  \citep*{pearl2018does}  admits that the choice of manipulation may in fact matter, but that this is simply not relevant for the definition of the \doo-operator, and that the causal effect is well-defined irrespective of this.  \citep*{pearl2018does}  discusses the third question from Section \ref{section:2.2}, the effect of obesity, and admits that the effect of changing obesity through exercise, diet or surgery may be very different, but that nevertheless the causal effect of obesity on mortality, corresponding to a {\it virtual intervention}, very much in the spirit of a strong version of the {\it ceteris paribus} condition,  is well defined:
\begin{quote}
``While it is true that the probability of death will generally
depend on whether we
manipulate obesity through diet versus, say, exercise, [...] ${\it do}({\rm obesity}=x)$ describes a virtual intervention, by which nature sets obesity to $x$, independent of diet or  exercise, while keeping every thing else intact, especially the processes that respond to $X$.'' (\citep*{pearl2018does}, p. 3).
\end{quote}
Although other researchers working on graphical models have taken a different view on the manipulability issues ({\it e.g.,} \citep*{richardson2013single, hernan2018data}), Pearl doubles down on his  position that tying causal effects to manipulations is not necessary in his most recent work:
\begin{quote} ``We end with the conclusion that researchers need not distinguish manipulable from non-manipulable variables; both types are equally eligible to receive the $\doo(x)$ operator'' (\citep*{pearl2019interpretation}, abstract).\end{quote}
I find the position that the manipulation is irrelevant  unsatisfactory, and find the insistence in the PO approach on a  theoretical or practical manipulation  helpful. I am not sure what is meant by ${\it do}({\rm obesity}=x)$ if the effect of changing obesity depends on the mechanism (say, exercise, diet, or surgery), and the mechanism is not specified in the operator. It is also not obvious to me why we would care about the value of
${\it do}({\rm obesity}=x)$  if the effect is not tied to an intervention we can envision. 
The insistence on manipulability in the PO framework resonates well in economics where policy relevance is a key goal ({\it e.g.,} \citep*{manski2013public}).
We are interested in policies that change the weight for currently obese people ({\it e.g.,} encouraging exercise, dietary changes, or surgery), or  that discourage currently non-obese people who are at risk of becoming obese from doing so (exercise,  dietary changes, or other life-style changes). What is relevant for policy makers is the causal effect of such policies, not the effect of a virtual intervention that makes currently obese people suddenly like non-obese people. In my original  1995 PhD  course with Don Rubin on causal inference we had a similar discussion about the ``causal effect of child poverty.'' From our perspective that question was ill-posed, and  a better-posed question would be about the effect of a particular intervention that would make currently poor families financially better off. One draconian policy to achieve this would be to take the children away from the poor families and put them with richer families. That does not seem like a policy the government should or would consider in this day and age. A more interesting policy would be to provide additional financial resources to currently poor families. The discusion in that class led to  a study of the effect of one such intervention, namely winning the lottery on subsequent outcomes (which ultimately led to \citep*{imbensrubinsacerdote}) as a way of illustrating the causal effect of a particular manipulation.

Without a specific manipulation in mind, it is also difficult to assess a particular identification strategy. It is only in the context of a particular treatment that it becomes clear whether in the analysis of the effects of obesity diet is a confounder or a mediator, partly dependent on the relative timing of the intervention. The flip side of that is that if the treatment is a virtual intervention, it becomes difficult to assess whether there are unobserved confounders.

As a third example to  clarify the benefits from being explicit about the nature of the manipulation, consider the statement: ``she did not get the position because she is a woman.'' In the PO approach such a statement is not clear without reference to some intervention. One such intervention could involve hiding the fact that the job candidate is a woman at the time of the interview from the individuals who make the hiring decisions. The recognition of the importance of being precise about the intervention is evident in \citep*{goldin2000orchestrating} who study the effect of blind auditions (behind curtains) for orchestras.

\subsection{Instrumental Variables, Compliers, and the Problem of Representing Substantive Knowledge}\label{section:iv}

In this section I  discuss instrumental variables and related methods from a PO and DAG perspective.  I want to make two main  points. First, some of the key assumptions in instrumental variables settings are not naturally captured in DAGs, whereas they are easily articulated in the PO framework. 
This extends to other shape restrictions that play an important  role in economic theory.
Second, one of the modern results in instrumental variables settings, the identification of the Local Average Treatment Effect (LATE, \citet{imbens1994, angrist1996}) is not easily derived in a DAG approach.\footnote{I should also note that the discussions of instrumental variables in the PO framework, and the subsequent Principle Stratification (PS) literature (\citet{frangakis2002principal, mealli2012refreshing}) shows that the claim in TBOW that ``the major assumption that potential outcome practitioners are invariably required to make is called `ignorability.' '' (TBOW, p. 281) is clearly incorrect. In the IV and PS settings a critical assumption is the exclusion restriction, similar to the absence of an arrow in a DAG, rather than an ignorability assumption.}

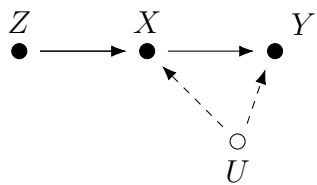
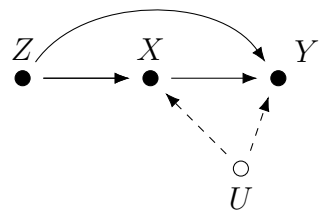
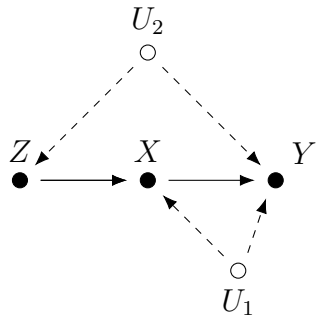
\begin{figure}
    \centering
    
    \begin{subfigure}[b]{0.45\textwidth}
    
    \begin{tikzpicture}[
        >=stealth,
        node distance=1.5cm        ]
        \node[observed, label=above:{\(Z\)}] (1) {};
        \node[observed, right=of 1, label=above:{\(X\)}] (2) {};
        \node[observed, right=of 2, label=above right:{\(Y\)}] (4) {};        
        \node[unobserved, below right=of 2, label=below:{\(U\)}] (5) {};
        \draw [->, notouch] (1.east) -- (2.west);
        \draw [->, notouch] (2.east) -- (4.west);
        \draw [->, notouch] (1.east) -- (2.west);
        \draw [dashed, ->, notouch] (5.north west) -- (2.south east);
        \draw [dashed, ->, notouch] (5.north east) -- (4.south west);
    \end{tikzpicture}    
        \caption{Instrumental Variables}
    \end{subfigure}
    \hfill
    \begin{subfigure}[b]{0.45\textwidth}

    \vspace{2em}
    \begin{tikzpicture}[
        >=stealth,
        node distance=1.5cm
        ]
        \node[observed, label=above:{\(Z\)}] (1) {};
        \node[observed, right=of 1, label=above:{\(X\)}] (2) {};
        \node[observed, right=of 2, label=above right:{\(Y\)}] (4) {};
        \node[unobserved, below right=of 2, label=below:{\(U\)}] (5) {};
        \draw [->, notouch] (1.east) -- (2.west);
        \draw [->, notouch] (2.east) -- (4.west);
        \draw [->, notouch] (1.east) -- (2.west);
        \draw [dashed, ->, notouch] (5.north west) -- (2.south east);
        \draw [dashed, ->, notouch] (5.north east) -- (4.south west);
        \draw [->, notouch] (1.north east) to [out=55, in=125] (4.north west);
    \end{tikzpicture}
        \caption{Violation of Exclusion Restriction in Instrumental Variables Setting}
    \end{subfigure}
    \hfill
    \begin{subfigure}[b]{0.45\textwidth}\vspace{2em}
    \begin{tikzpicture}[
        >=stealth,
        node distance=1.5cm        ]
        \node[observed, label=above:{\(Z\)}] (1) {};
        \node[observed, right=of 1, label=above:{\(X\)}] (2) {};
        \node[observed, right=of 2, label=above right:{\(Y\)}] (4) {};        
        \node[unobserved, below right=of 2, label=below:{\(U_1\)}] (5) {};
        \node[unobserved, above=of 2, label=above:{\(U_2\)}] (6) {};
        \draw [->, notouch] (1.east) -- (2.west);
        \draw [->, notouch] (2.east) -- (4.west);
        \draw [->, notouch] (1.east) -- (2.west);
        \draw [dashed, ->, notouch] (5.north west) -- (2.south east);
        \draw [dashed, ->, notouch] (5.north east) -- (4.south west);
        \draw [dashed, ->, notouch] (6.south west) -- (1.north east);
        \draw [dashed, ->, notouch] (6.south east) -- (4.north west);
    \end{tikzpicture}
        \caption{Violation of Exogeniety Assumption in Instrumental Variables Setting}
    \end{subfigure}
  \caption{\label{fig_iv2} Instrumental Variables}
\end{figure}

Instrumental variables have played an important role in econometrics since the 1920s. 
Whereas causal inference in statistics started with randomized experiments, econometricians were more interested in settings where the assignment mechanism reflected choice rather than chance (\citep{imbens2014}).
They continue 
to be an important topic for current research in various nonparametric settings, {\it e.g.,}
\citep*{horowitz2011applied, imbens2009identification, d2015identification, torgovitsky2015identification}. Here I focus on the simplest case with a single binary instrument.
Let me start by recalling the DAG for a simple instrumental variables setting, in Figure \ref{fig_iv2}(a). 
Endogeneity is captured by the presence of arrows between the unobserved confounder $U$ and the treatment $X$ and the outcome $Y$. In economics the endogeneity often arises from agents actively making choices regarding the causal variable on the basis of anticipated effects of those choices (\citep*{athey1998empirical, imbens2014}).
The first of the  two key assumptions is that there are no direct effect of the instrument $Z$ on the outcome $Y$. If such a direct effect were present the DAG would look like Figure \ref{fig_iv2}(b).
 The second key assumption is that there is no unmeasured confounder that affects the instrument and  the outcome $Y$. The DAG in Figure \ref{fig_iv2}(c) has such an unmeasured confounder.
In the PO approach in \citet{angrist1996} 
the starting point is the postulation of the potential outcomes $Y_i(z,x)$, indexed by both the treatment and the instrument, and the potential treatment status $X_i(z)$, indexed by the instrument. 
The first assumption is the exclusion restriction that the potential outcomes do not vary with the instrument,  $Y_i(z,x)=Y_i(z',x)$ for all $z,z'$, and the second is an unconfoundedness assumption,
$Z_i\indep (Y_i(0,0),Y_i(0,1),Y_i(1,0),Y_i(1,1),X_i(0),X_i(1))$.
Again one may well find the graphical representation of these assumptions more attractive or accessible than the algebraic representation in the PO framework, or the other way around.

A third key assumption in the instrumental variables setting is the monotonicity condition that the instrument has a monotone effect on the treatment. In the PO formulation the assumption is $X_i(1)\geq X_i(0)$, also referred to as the no-defiance assumption (\citet{imbens1994, baker1994paired, angrist1996, baker2016latent}). This assumption captures the notion that the instrument is an incentive. Although agents need not respond to such incentives ($X_i(1)$ may be equal to $X_i(0)$ for some individuals), they do not respond perversely to them (there are no individuals with $X_i(1)<X_i(0)$, the so-called defiers).
 In economics much information comes in the form of such monotonicity, or more generally, shape restrictions. Utility functions are typically at least weakly increasing  in their arguments ({\it e.g.,} quantities of goods, leisure) over the relevant ranges, as well as concave. Combined with budget constraints this leads to demand functions being decreasing in prices. By the same argument, instruments that can be interpreted as changing incentives lead to responses that are monotone in these instruments.
Because the compliance types are defined explicitly in terms of potential outcomes, this condition is easy to articular in the PO framework.
 DAGs do not naturally represent such conditions, as was pointed out
in \citep*{imbens1995causal} in a comment on \citep*{pearl1995causal}.  Twenty-five years later this has not been addressed, although that may change:
\begin{quote} ``Judging from my discussions with @eliasbareinboim and @analisereal, I think that incorporating shape restrictions such as monotonicity into the [DAG] framework, to see how it creates opportunities for identification, will be high on the agenda of computer scientists in the coming years.''
(April 4th, 2019, Twitter,  Paul H\'unermund, @PHuenermund)
\end{quote}
There are already some attemps to include these assumptions in the DAGs, {\it e.g.,}  \citep*{steiner2017graphical}, but I do not find  these DAGs  particularly illuminating.

Given the assumptions, the identification results are also easier to derive in the PO framework.
The early results in the instrumental variables literature with heterogenous treatment effects  established that the average effect of the treatment was not identified in general (\citet{heckman1990varieties}). In addition  bounds were derived 
(\citep*{robins1986new, manski_bounds, balke1997bounds, swanson2018partial}.
Such results follow easily in both the DAG and PO approaches. However, the subsequent identification result for the average effect for compliers, the Local Average Treatment Effect (LATE),
\[ \tau^{\rm late}=\mathbb{E}[Y_i(1)-Y_i(0)|X_i(0)=0,X_i(1)=1],\]
 is more difficult to derive using a graphical approach. In fact, setting the problem up in the PO framework led (\citep{imbens1994, angrist1996})  to their result. The PO framework  is a natural one here because the very definition of the LATE estimand involves the potential outcomes $X_i(0)$ and $X_i(1)$ that define compliance status.
In passing TBOW acknowledges the difficulty the DAG approach has in this setting:
\begin{quote}``To sum up, instrumental variables are an important tool in that they help us uncover causal information that goes beyond the {\it do}-calculus. [...]  if we can justify an assumption like monotonicity ... on scientific grounds, then a more special-purpose tool like instrumental variables is worth considering.'' (TBOW, p.257)\end{quote}
This reflects more general difficulties in the DAG framework to capture individual level heterogeneity 
(\citep*{hartman2015sate, wager2017estimation}).

TBOW also discusses  instrumental variables in the context of one of the classic examples, non-compliance in a randomized experiment ({\it e.g.,} \citet{imbens1997bayesian}). Here the instrument is the random assignment  to either the cholesterol drug or a placebo. The treatment of interest or endogenous variable is the receipt of the cholesterol drug. The outcome is the subsequent cholesterol level. In this case the instrument is unconfounded by design: the randomization rules out the presence of the unmeasured confounders in Figure   \ref{fig_iv2}(c). However, the exclusion restriction that rules out the direct effect in Figure  \ref{fig_iv2}(b) is still controversial. Although TBOW blithly dismisses concerns with the exclusion restriction in this setting  by writing that  ``Common sense suggests that there is no way that receiving a particular random number $(Z)$ would affect cholesterol $(Y)$'' (TBOW, p. 253), there may in fact well be violations of the exclusion restriction. For example, side-effects of the actual drug  could lead the patients to non-comply, but seek other treatments. 
Simply leaving out an arrow because of ``common sense'' may not be sufficient to make identifying assumptions.
The careful discussion of such possible violations of the exclusion restriction from a PO perspective is a hallmark of modern empirical economic practice, through sensitivity analyses ({\it e.g.,} \citet{angrist1996}) and other supplementary analyses (\citet{athey2017state,
ding2017instrumental}). 
For example, in the seminal \citet{angrist1990lifetime} instrumenal variables application it is helpful to consider separately the exclusion restriction (the absence of a direct effect of the instrument, the draft lottery number, on the outcome, earnings) for never-takers and always-takers, because it may be much more plausible for some groups than for others.
Such discussions underline  how rare it is in practice to have settings where the absence of arrows is immediately credible.

\begin{figure}
\begin{subfigure}[b]{0.45\textwidth}
\centering
\begin{tikzpicture}
        \draw[->] (-0.5,0) -- (4,0) node[right] {$x$};
        \draw[->] (0,-0.5) -- (0,2) node[above] {$y$};

        \draw[dashed,-] (2,-0.1) node[below]{c} -- (2,1.75);

        \draw[domain=0.1:2,smooth,variable=\x] 
            plot ({\x},{0.5 + 0.1 * sin(deg(5 * \x + 1.5)) + 0.1 * \x});
        \draw[domain=2:3.9,smooth,variable=\x] 
            plot ({\x},{0.5 + 0.5 + 0.1 * sin(deg(5 * \x + 1.5)) + 0.1 * \x});

        \node (1) at (5,1.5) {\(E(y|x)\)};
\end{tikzpicture}
\caption{\label{ce} Conditional Expectation}
\end{subfigure}
\hfill
\begin{subfigure}[b]{0.45\textwidth}
\centering
\begin{tikzpicture}
        
        \def\mu{2.5}  
        \def\s{1}  
        \def\a{3.25}  
        \def\d{0.2}  
        \draw[->] (-0.5,0) -- (4,0) node[right] {$x$};
        \draw[->] (0,-0.5) -- (0,2) node[above] {$f(x)$};

        \draw[dashed,-] (2,-0.1) node[below]{c} -- (2,1.75);

        \draw[domain=0.1:1.5,smooth,variable=\x] 
            plot ({\x},{\a*exp(-(\x-\mu)^2/\s^2)/sqrt(2 * pi * \s^2)});
        \draw[domain=1.5:2,smooth,variable=\x] 
            plot ({\x},{\a*(-\d*(1.5-\x)^2 + exp(-(\x-\mu)^2/\s^2)/sqrt(2 * pi * \s^2))});
        \draw[domain=2:2.5,smooth,variable=\x] 
            plot ({\x},{\a*(\d*(2.5-\x)^2 + exp(-(\x-\mu)^2/\s^2)/sqrt(2 * pi * \s^2))});
        \draw[domain=2.5:3.9,smooth,variable=\x] 
            plot ({\x},{\a*exp(-(\x-\mu)^2/\s^2)/sqrt(2 * pi * \s^2)});

\end{tikzpicture}
\caption{\label{density} Density}
\end{subfigure}
\caption{\label{fig_rd}  Regression Discontinuity}
\end{figure}
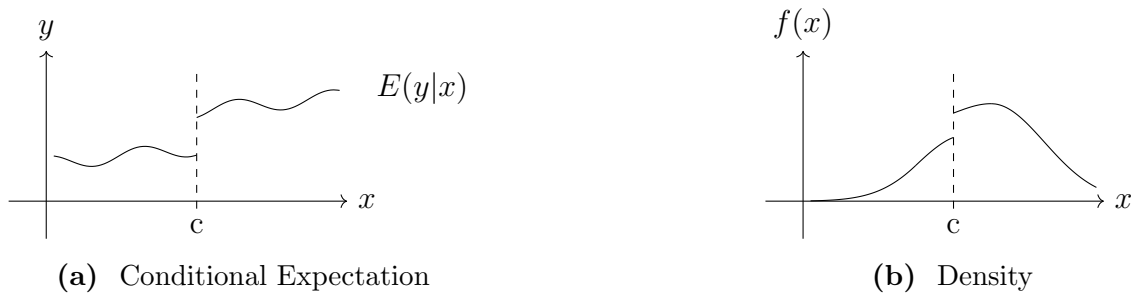

A closely related setting where DAGs have also not added much insight is in {\it regression discontinuity designs} (RDDs) (\citet{campbell, imbenslemieux, calonico, lee2010regression}). In the {\it sharp} RDD the assignment rule for the binary treatment $X$ is a deterministic function of a running pretreatment variable $W$:
\[ X(w)={\bf 1}_{w\leq c},\]
for some fixed threshold $c$. The average causal effect of the treatment at the threshold is identified in that case under smoothness conditions on the expected values or the distribution of the potential outcomes conditional on the running variable. 
A long time after RD designs were first introduced in \citep*{campbell}, they have become a very popular identification strategy in economics with many credible applications. 
See \citep*{cook} for a historical perspective.
One reason is that there are many settings where it is attractive for administrators to set fixed assignment rules that justify such identification strategies, although it does not merit any discussion in TBOW.
Although DAGs have been proposed
for this setting, \citep*{steiner2017graphical}, I do not think they  clarify the identification results, 
where the critical assumptions include a discontinuity in one conditional expectation, and smoothness of other conditional expectations.
Nor do the DAGs illuminate any of the issues that have occupied the methodological literature in this area. Such issues include violations of the smoothness assumption because of manipulation of the running variable, which motivated the McCrary test (\citet{mccrary}).
Other concerns  relate to estimation issues stemming from the problem of estimating a regression function at a single point (\citet{imbenskalyanaraman, calonico, armstrong2018optimal, imbens2018optimized}).
Although DAGs are rarely seen in regression discontinuity analyses, it is not because figures are not viewed as useful there.
Figure \ref{fig_rd}(\subref{ce}) and \ref{fig_rd}(\subref{density})  are typical of the figures that are routinely presented in regression discontinuity analyses, and more than any DAG for this case, it clarifies what is going on here. There is a running variable $X$, and at a particular threshold  $c$ the conditional mean of the outcome changes discontinuously as in Figure  \ref{fig_rd}(\subref{ce}). This jump is interpreted as the causal effect of the change in the participation rate. Second, there is a concern that the running variable may have been manipulated. Such violations would show up in a discontinuity in the density of the running variable at the threshold assessed in the McCrary test and illustrated  in \ref{fig_rd}(\subref{density}).

\subsection{Simultaneity}

DAGs are by their very definition not cyclical, and as such do not naturally capture assumptions about equilibrium behavior, although there recently is some work going in that direction ({\it e.g., } \citep*{forre2019causal}). Equilibrium assumptions are of course central to economics. Identification and estimation of supply and demand functions in competitive markets is at the core of the econometrics, and has been since the early days in econometrics in the 1930s and 1940s ({\it e.g.,} \citep*{tinbergen1930determination, haavelmo1943statistical}).
In a PO framework equilibrium notions are accomodated very naturally.  Let the demand function and supply function in market $t$ be
\[ q^d_t(p,x^d),\hskip1cm {\rm and}\  \ q^s_t(p,x^s).\]
Here $x^d$ and $x^s$ are exogenous demand and supply shifters, with the demand shifters only entering the demand function, and the supply shifters only entering the supply function.
\citep*{haavelmo1943statistical} and \citep*{tinbergen1930determination} are very clear  that these potential outcome functions are well defined for all values of the prices and shifters, not just the realized ones.
 Given the values of the shifters in market $t$, $x^d_t$ and $x^s_t$ respectively, the equilibrium price that we observe in this market is the price $p_t$ that solves
\[p_t\ \ {\rm solves}\ \  q^d(p,x^d_t)=q^s_t(p,x^s_t),\]
and the observed quantity traded is $q_t=q^d(p_t,x^d_t)=q^s_t(p_t,x^s_t)$.

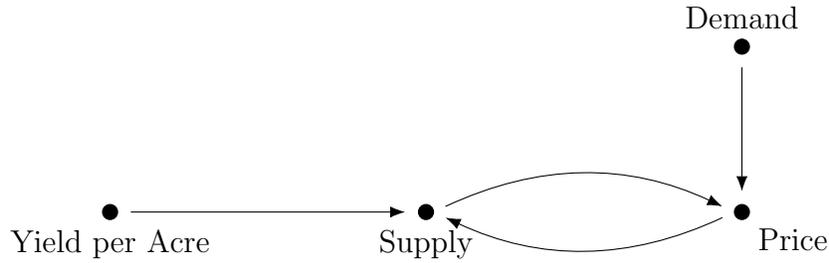
\begin{figure}

    \begin{tikzpicture}[
        >=stealth,
        node distance=2cm and 4cm
        ]

        \node[observed, label=below:{Yield per Acre}] (1) {};
        \node[observed, right=of 1, label=below:{Supply}] (2) {};
        \node[observed, right=of 2, label=below right:{Price}] (3) {};
        \node[observed, above=of 3, label=above:{Demand}] (4) {};

        \draw [->, notouch] (1.east) -- (2.west);

        \draw [->, notouch] (2.east) to [out=25, in=155] (3.west);
        \draw [->, notouch] (3.west) to [out=-155, in=-25] (2.east);

        \draw [->, notouch] (4.south) -- (3.north);

    \end{tikzpicture}

\caption{\label{fig_pearl_simeq} Based on Figure 7.10 in TBOW, p. 251.}
\end{figure}

TBOW touches on the simultaneous equations case briefly, but dismisses any real concerns with simultaneity:
\begin{quote} ``Figure 7.10 [reproduced as Figure \ref{fig_pearl_simeq} here, GWI] shows a somewhat simplified version of Wright's diagram. Unlike most diagrams in this book, this one has  `two-way' arrows, but I would ask the reader not to lose too much sleep over it. With some mathematical trickery we could equally well replace the Demand$\rightarrow$Price$\rightarrow$Supply chain with a single arrow Demand$\rightarrow$Supply, and the figure would then look like Figure 7.9 (though it would be less acceptable to economists).''
(TBOW, p. 250-251).
\end{quote}
Here Pearl's lack of engagement with the data shows. It is not clear what is  meant by the variables ``Supply'' and ``Demand'' that may affect each other, and in particular how these variables relate to the variables we typically think of observing in such settings, namely prices, quantities, and possibly some exogenous demand and supply shifters (a second question is  why there is an asymmetry between Suppy and Demand, where Demand has an arrow going into Price, but Supply has both an arrow going into Price and an arrow coming from Price). 
In 1998 Pearl invited me to present  the supply and demand analysis in  \citep*{angrist2000interpretation} to his group at UCLA. Unfortunately my presentation seems not to have succeeded in giving the audience a full understanding of the way economists think about supply and demand, although it seems to have made Pearl  aware that his diagram 7.10 (Figure \ref{fig_pearl_simeq})  may not be fully capturing the way economists do think about these problems.

\begin{figure}
    \centering
    
    \begin{subfigure}[b]{0.45\textwidth}
    
    \begin{tikzpicture}[
        >=stealth,
        node distance=1cm and 1cm
        ]
        \node[observed, label=left:{\(x^d\)}] (1) {};
        \node[observed, above right=of 1, label=above left:{\(q\)}] (2) {};
        \begin{scope}[node distance=1cm and 2cm]
        \node[observed, right=of 2, label=above right:{\(p\)}] (3) {};
        \end{scope}
        \node[observed, below right=of 3, label=right:{\(x^s\)}] (4) {};
        \draw [->, notouch] (1.north east) -- (2.south west);
        \draw [->, notouch] (4.north west) -- (3.south east);
        \draw [->, notouch] (2.north east) to [out=60, in=120, looseness=1] node[above]{demand} (3.north west);
        \draw [->, notouch] (3.south west) to [out=-120, in=-60, looseness=1] node[below]{supply} (2.south east);
    \end{tikzpicture}
            \caption{Demand and Supply I}
    \end{subfigure}
    \hfill
    \begin{subfigure}[b]{0.45\textwidth}   
    \vspace{2em}
    \begin{tikzpicture}[
        >=stealth,
        node distance=1cm and 1.4cm
        ]
        \node[observed, label=left:{\(x^d\)}] (1) {};
        \node[observed,  right=of 1, label=above:{\(q^d(\cdot)\)}] (2) {};
        \begin{scope}[node distance=1cm and 2cm]
        \node[observed,  right=of 2, label=above right:{\(q^s(\cdot)\)}] (3) {};
        \end{scope}
        \node[observed, right=of 3, label=right:{\(x^s\)}] (4) {};
        \node[observed, above right=of 2, label=above:{\(p^e\)}] (5) {};
        \node[observed, below right=of 2, label=below:{\(q^e\)}] (6) {};
        \draw [->, notouch] (1.east) -- (2.west);
        \draw [->, notouch] (4.west) -- (3.east);
        \draw [->, notouch] (2.north east) -- (5.south west);
        \draw [->, notouch] (3.north west) -- (5.south east);
        \draw [->, notouch] (2.south east) -- (6.north west);
        \draw [->, notouch] (3.south west) -- (6.north east);
    \end{tikzpicture}  
        \caption{Demand and Supply II}
    \end{subfigure}
  \caption{\label{fig_simeq} Simultaneous Equations}
\end{figure}
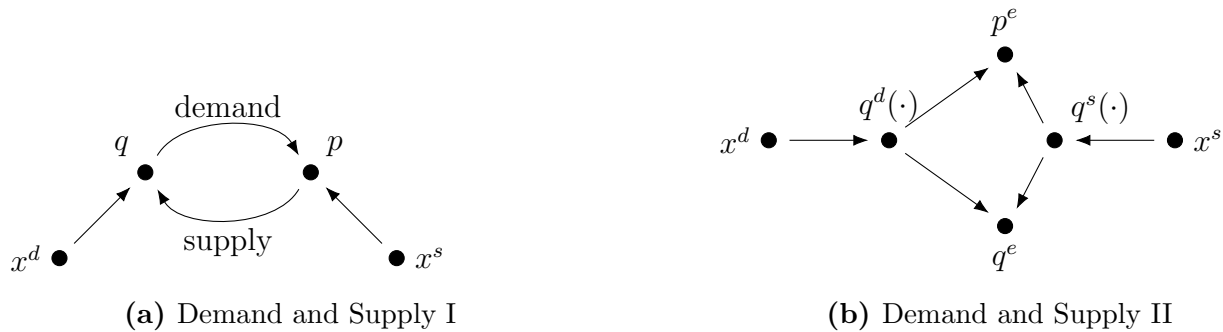

In fact it is not immediately obvious to me how one would capture supply and demand models in a  DAG. In Figures \ref{fig_simeq}(a) and \ref{fig_simeq}(b) there are two attempts I made. The first, Figure \ref{fig_simeq}(a), captures the simultaneity by having arrows go from prices to quantities and back. This does not, in my view, really capture what is going on. It is not the case that there is an effect of prices on quantities that corresponds to the demand function and an effect from quantities to prices that captures supply. Rather, as in Figure \ref{fig_simeq}(b), the demand and supply function are the primitives. Together they determine both equilibrium prices and quantities. The identification comes from the presence of  observed demand and supply shifters. However, in this figure the nodes $q^d(\cdot)$ and $q^s(\cdot)$ are not variables, but functions, so it is not clear that the graphical machinery such as the \doo-calculus can illuminate the identification issues.
It would be helpful if the DAG researchers engaged more fully with such models where equilibrium plays a central role.
See also 
\citep{richardson2014ace} for discussions of modeling equilibrium conditions from a non-economic perspective.

\subsection{Unconfoundedness, The Choice of Conditioning Variables and M-Bias}

The unconfoundedness setting where the researcher adjusts for some variables in order to estimate the causal effect of a binary treatment  on some outcome is probably the most important one in practice in the modern CI literature. There is much theoretical statistical theory developed  for this case and it underlies much empirical work in economics.
In the PO literature the starting point is the assumption that conditional on the observed confounders $W_i$ the cause $X_i$ is independent of the potential outcomes $Y_i(0)$ and $Y_i(1)$, or in the standard notation,
\begin{equation}\label{unconf} X_i\  \indep\ \Bigl(Y_i(0),Y_i(1)\Bigr)\ \Bigl|\ W_i.\end{equation}
A classic example is the Lalonde study (\citet{lalonde}), which has been re-analyzed many times ({\it e.g., } \citet{heckmanhotz, dehejiawahba, abadie2011bias}), and which currently is a standard data set on which to try out new methods (see \citep{athey2019using} for a discussion regarding simulation studies in this context). In this study the treatment of interest is a labor market training program, the outcome is earnings, and the eight pre-treatment variables include measures of educational achievement, ethnicity, marital status, age, and  prior measures of annual earnings. One DAG corresponding to this is Figure 8(a). We could also include any number of arrows between the pre-treatment variable without changing the identification strategy based on adjusting for the full set of pre-treatment variables.

One may wish to argue whether the DAG here expresses the content of the critical assumption more clearly than the conditional independence statement in (\ref{unconf}). \citep*{pearl2012causal} takes a strong view on this:
\begin{quote}
``The weakness of this [the potential outcome, GWI] approach surfaces in the problem formulation phase where, deprived of diagrams and structural equations, researchers are forced to express the (inescapable)
assumption set A in a language totally removed from scientific knowledge, for example, in the form of conditional independencies among counterfactual variables'' (\citep*{pearl2012causal} p. 16-17).
\end{quote}
I think that statement misses the point. This setting, where the critical assumption is ignorability or unconfoundedness, is so common and well studied, that merely referring to  its label is probably sufficient for researchers to understand what is being assumed. Adding a DAG, or for that matter adding a proof that the average causal effect is identified in that setting, is superfluous because researchers are familiar with the setting and its implications.

One question that needs to be answered prior to the choice of estimation methods,  is the choice of variables $W_i$ to condition on.
In the PO literature it is typically taken as given which variables $W_i$ need to be conditioned on.
Often it is mentioned that a requirement is that these are {\it pre-treatment variables}, that is, variables that are not affected themselves by the treatment because they assume their value prior to the determination of the treatment. 
\citet{rosenbaum1984consequences} warns about the dangers of adjusting for variables that are not proper pre-treatment variables. Although one can generate examples where it would be valid to do so, there do not appear to be many realistic applications where adjusting for post-treatment variables as if they are pre-treatment variables is helpful. In practice the confusion often concerns what the treatment is and what that implies for the classification as pre- or post-treatment variables.
However, if the variables under consideration are viewed as proper pre-treatment variables the recommendation in the PO literature is typically to include all of them, possibly subject to concerns about efficiency (adding variables that have no association with the potential outcomes could lower precision). There are some exceptions to this. In settings where some of the pre-treatment variables are  instruments, researchers typically follow different strategies. In practice I have not seen much evidence of confusion in such cases.

\begin{figure}
    \centering
    \begin{subfigure}[b]{0.45\textwidth}
    \begin{tikzpicture}[
        >=stealth,
        node distance=1.2cm and 2cm
        ]
        \node[unobserved, label={[align=center]above:attitude towards \\ social norms}](1) {};
        \begin{scope}[node distance= 3cm]
        \node[observed, below=of 1, label=below:{smoking}] (2) {};
        \end{scope}
        \node[observed, below right=of 1, label={[align=center]below:seat belt \\ use}] (3) {};
        \node[unobserved, above right=of 3, label={[align=center]above:attitudes towards\\safety and health\\related measures}] (4) {};
        \begin{scope}[node distance= 3cm]
        \node[observed, below=of 4, label=below:{lung cancer}] (5) {};
        \end{scope}
        \draw [dashed, ->, notouch] (1.south) -- (2.north);
        \draw [dashed, ->, notouch] (1.south east) -- (3.north west);
        \draw [dashed, ->, notouch] (4.south west) -- (3.north east);
        \draw [dashed, ->, notouch] (4.south) -- (5.north);
         \draw [ ->, notouch] (2.west) -- (5.east);
    \end{tikzpicture}    
        \caption{\label{bias1} M-Bias Assumption Satisfied}
    \end{subfigure}
    \hfill
    \begin{subfigure}[b]{0.45\textwidth}   
    \vspace{2em}
    \begin{tikzpicture}[
        >=stealth,
        node distance=1.2cm and 2cm
        ]
        \node[unobserved, label={[align=center]above:attitude towards \\ social norms}](1) {};
        \begin{scope}[node distance= 3cm]
        \node[observed, below=of 1, label=below:{smoking}] (2) {};
        \end{scope}
        \node[observed, below right=of 1, label={[align=center]below:seat belt \\ use}] (3) {};
        \node[unobserved, above right=of 3, label={[align=center]above:attitudes towards\\safety and health\\related measures}] (4) {};
        \begin{scope}[node distance= 3cm]
        \node[observed, below=of 4, label=below:{lung cancer}] (5) {};
        \end{scope}
        \draw [dashed, ->, notouch] (1.south) -- (2.north);
        \draw [dashed, ->, notouch] (1.south east) -- (3.north west);
        \draw [dashed, ->, notouch] (4.south west) -- (3.north east);
        \draw [dashed, ->, notouch] (4.south) -- (5.north);
        \draw [dashed, ->, notouch] (4.south west) to [out=-105, in=5] (2.north east);
  \draw [ ->, notouch] (2.west) -- (5.east);
        \end{tikzpicture}
       \caption{\label{bias2} Violation of M-Bias Assumption}
    \end{subfigure}
  \caption{\label{fig_mbias} M-Bias}
\end{figure}

The DAG literature is more concerned with the systematic  choice of conditioning variables, even in settings where all of these are proper pre-treatment variables (\citep{vanderweele2011new, rotnitzky2019efficient, haggstrom2018data}). A key example used  in TBOW and 
\citet{pearl2009myth}, exhibits what is called  M-bias. Consider Figure \ref{fig_mbias}(\subref{bias1}). Using the example in TBOW, the interest is in the causal effect of smoking  on lung cancer. All three of the remaining variables are proper pre-treatment variables. Suppose that only seat belt use is observed, with the other two pre-treatment variables, attitudes towards social norms and safety and health measures  unobserved. Note that in general  seat belt use  will be correlated with both smoking and lung cancer in the DAG in Figure \ref{fig_mbias}(\subref{bias1}) because of the back-door paths from seat belt use to smoking and lung cancer. There is only one backdoor path from smoking to lung cancer, namely (smoking $\rightarrow$ social norms $\rightarrow$ seat belt $\leftarrow$ safety/health $\rightarrow$ lung cancer). Because seat belt usage is a collider on this path there is no need to condition on any variable, and without any conditioning we can identify the causal effect of smoking  on lung cancer because $P($lung cancer$|\doo($smoking$))=P($lung cancer$|$smoking). However, again because seat belt usage is a collider on this backdoor path, conditioning on pre-treatment variable  seat belt usage, would introduce bias. (We would have no bias if we conditioned on the other two pre-treatment variables, social norms and safety/healt attitudes, but because these are not observed this is not feasible.) This all follows directly from the back-door criterion, as pointed out in TBOW.
In TBOW and \citet{pearl2009myth} this example comes up repeatedly as a warning against the common practice in the PO literature to adjust for any pre-treatment variables that are correlated with the treatment and the outcome.

This is an interesting case. There is no doubt that if we are confident that this is, at least approximately, the appropriate DAG, we should not condition on seat belt use. However, it is not clear how relevant this case is for empirical practice. The main example that is given in TBOW and  \citet{pearl2009myth} is motivated by an analysis carried out by Donald Rubin.
Let us consider the DAG in Figure \ref{fig_mbias}(\subref{bias1}) for a moment.
First of all, seatbelt usage may in fact not be a pre-treatment variable because it may reflect behavior after smoking was started. I will put that issue aside for the moment.
 It seems reasonable to assume there is no direct causal effect of seatbelts on either smoking or lung cancer, so I am comfortable with no arrows from seat belt use to either smoking  or lung cancer. However, the implicit assumption that there is no effect of attitudes towards safety and health-related measures on smoking seems very implausible.
 In all the discussions of this example, Pearl never discusses whether this assumption is plausible. In order to use these methods, one needs to carefully consider every absent link, and in a setting with as many variables as there are in Figure \ref{fig_mbias}(\subref{bias2}) that is a daunting task.
  Now suppose we allow for one such direct effect, leading to the DAG in Figure \ref{fig_mbias}(\subref{bias2}), or alternatively with an arrow from attitudes towards social norms to cancer. Now the causal effect of smoking on lung cancer is not identified. Conditioning on seat belt use introduces a new bias, but not adjusting for differences in seat belt use between smokers and non-smokers leaves a bias. What to do? 
  Here seat belt use is essentially a proxy for the unobserved attitude variable, and whether to use it is a difficult question that does not have a single right anwer.
  The researcher has to make a choice which of the two evils is the lesser one, or possibly do a sensivity analysis. I am not sure whether adjusting for seat belt usage between smokers and non-smokers, as Rubin did, is a good thing or not, but certainly I would not rule out doing this. 
The recent analysis in   \citep*{ding2015adjust} suggests that, both in this specific example, and more generally when there is concern with confounding as well as M-bias, adjusting may still be superior.
Beyond that, I want to point out that that, as in the front-door criterion setting, even in the examples TBOW chooses to discuss, it is not at all clear that the method that TBOW favors actually works.
  This is particularly important because TBOW introduces this example after admitting that there was concern that such a DAG might be very implausible:
 \begin{quote}``When I started showing this diagram to statisticians in the 1990s, some of them laughed it off and said that such a diagram was extremely unlikely to occur in practice. I disagree!''
(TBOW, p. 161) \end{quote}

 Another way to make the point that this may be an unusual setting in practice, note that if the DAG in Figure \ref{fig_mbias}(a) is correct, it means that we can identify the causal effect of smoking on lung cancer, but we cannot identify the effect  for those who wear seat belts, because there is a hidden confounder there, nor can we identify the effect  for those not wearing seat belts, because there is a hidden confounder there. Mixing those who wear seat belts and those who do not, exactly cancels the two biases and leaves us with a valid casusal effect. This does appear an ``extremely unlikely'' setting.

To conclude this discussion, let us return to the Lalonde study. The M-bias argument would suggest that if we carefully model the causal relationships between all eight of the pre-treatment variables, we might come up with a structure that could suggest that we should not adjust for differences in some of these pre-treatment variables. It appears difficult to me to find a credible story that would lead to such a conclusion. In the end  a DAG like Figure \ref{fig_unc_many}(a), possibly augmented with some arrows between the pre-treatment variables would be just as plausible as any such alternative.

\subsection{Counterfactuals}

Counterfactuals occupy the third and highest rung of the ladder of causality in TBOW. The identification of counterfactuals is a very tricky issue, and I want to clarify some of the issues that are raised in TBOW and compare the answers to what a PO-based approach might say. On page 273 TBOW presents an example of such a question: Alice has a high school degree, 6 years of experience, and her annual earnings are \$81K. TBOW asks: ``What would Alice's salary be if she had a  college degree?'' (TBOW, p. 273).  
This is a difficult question. It comes up in other settings as well where it is generally  fraught with complications. Suppose someone took a drug, and they died. Would they not have died had they not taken the drug? A randomized experiment can only identify the fraction of survivors under both taking the drug and not taking the drug, within subpopulations defined in terms of pre-treatment characteristics. Such an experiment cannot identify {\it which} individuals would have survived under the alternative treatment, without strong additional assumptions.

\begin{figure}
    \vspace{0.5em}
    \begin{tikzpicture}[
        >=stealth,
        node distance=3cm
        ]
        \node[observed, label=below:{\(Education\)}] (1) {};
        \node[observed, right=of 1,  label=below:{\(Salary\)}] (2) {};
        \node[observed, above right=of 1, label=above:{\(Experience\)}] (3) {};

        \draw [->, notouch] (1.east) -- (2.west);
        \draw [->, notouch] (1.north west) -- (3.south east);
        \draw [->, notouch] (3.north west) -- (2.south west);
    \end{tikzpicture}\vspace{3em}
    
    \caption{\label{fig_alice} Based on Figure 8.3, TBOW, p. 276}
   \end{figure}
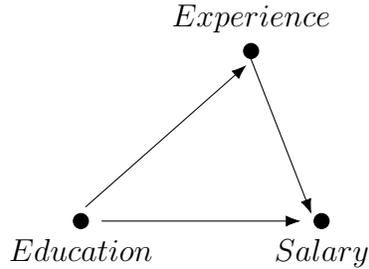

So what do the authors do here?
To answer the question about Alice's salary had she had a college degree, given that in fact she had a high school degree, TBOW presents the DAG in Figure \ref{fig_alice}. 
The authors then go further, and set up a {\it Structural Causal Model}, where they specify salary $(Y)$ as a function of education ($X)$ and experience $(W)$ and an unobserved component, and similarly for experience as a function of education and an unobserved component as
\[ Y=f_Y(X,W,U_Y),\hskip1cm {\rm and} \  W=f_W(X,U_W).\]
They then simplify things ``assuming linear functions throughout'' (TBOW, p. 277):
\[ Y=\alpha_0+\alpha_1 X+\alpha_2 W+U_Y,\hskip1cm
 W=\beta_0+\beta_1 X+U_W.\]
 This allows them to first estimate the  coefficients  $(\alpha_0,\alpha_1,\beta_0,\beta_1,\beta_2)$ and then, importantly, the unobserved components $U_Y$ and $U_W$ for Alice as 
 \[  U_{Y,{\rm Alice}}=Y_{\rm Alice}-\alpha_0-\alpha_1 X_{\rm Alice}-\alpha_2 W_{\rm Alice},\]
 \[  U_{W,{\rm Alice}}=X_{\rm Alice}-\beta_0-\beta_1 X_{\rm Alice}.\]
   Given these coefficients and residuals TBOW then estimates Alice's salary
 given a college degree $(X_{\rm college}$) as
 \[ \hat Y_{\rm Alice}=\alpha_0+\alpha_1 X_{\rm college}+\alpha_2 \Bigl(\beta_0+\beta_1 X_{\rm college}+U_{W,{\rm Alice}}\Bigr)+U_{Y,{\rm Alice}}.\]

From a PO perspective a major question concerns the way the residuals $ U_{Y,{\rm Alice}}$ and
$ U_{W,{\rm Alice}}$ are used. Starting with potential outcomes, $W_{\rm Alice}({\rm highschool})$ and $W_{\rm Alice}({\rm college})$, 
the residuals would be also be indexed by the treatment:
\[ U_{W,{\rm Alice}}({\rm highschool})=W_{\rm Alice}({\rm highschool})-\beta_0-\beta_1{\rm highschool},\]
and
\[ U_{W,{\rm Alice}}({\rm college})=W_{\rm Alice}({\rm college})-\beta_0-\beta_1{\rm college}.\]
 There is often a reluctance to rely on assumptions about the association between the residuals $U_{W,{\rm Alice}}({\rm highschool})$ and $U_{W,{\rm Alice}}({\rm college})$ from potential outcomes after conditioning on other variables, whereas the discussion in TBOW implicitly assumes the two are equal.
These issues involving unit-level heterogeneity come up in many settings, {\it e.g,} \citep*{sekhon2017inference}.
Putting aside for a moment the other variables, and assume we actually have a randomized experiment. 
We would not be confident that the  rank of $Y_{\rm Alice}({\rm highschool})$ in the distribution of high-school salaries would be the same as the rank of $Y_{\rm Alice}({\rm college})$ in the distribution of college salaries (corresponding in the linear case to equality of the residuals $U_{W,{\rm Alice}}({\rm highschool})$ and $U_{W,{\rm Alice}}({\rm college})$). 
There is obviously many reasons why an individual who would be relatively productive given a college degree ({\it e.g.,} someone with strong cognitive abilities), would not necessarily be relatively productive with only a high-school degree. The jobs a person would end up with given a college degree would be different from the one they would have with a high-school degree, with value placed on different sets of skills. This type of comparative advantage argument is common in economics, and it is a critical part of the Roy model of occupational choice (\citep*{roy1951some}).
TBOWs implicit assumption of a single (scalar) unobserved component is quite distinct from the assumption of linearity of the conditional expectations that TBOW bundles it with, but it is crucial to the ability  to predict Alice's salary given a college degree. 

This is not to say that this type of assumption is never made in economics. The point is that when it is done well, such assumptions are  typically  made very explicitly, with full recognition of the identifying power it has ({\it e.g.,} the discussion of the rank correlation assumption in \citep*{chernozhukov2005iv}), rather than just as one component  of a set of functional form assumptions the way TBOW does here. It has also received much attention in the discussion on identifying distributions of treatment effects, as opposed to differences in the distribution of potential outcomes (see the discussion in
\citep*{manski1996learning, deaton2010, imbens2010}). The way in which TBOW exploits this assumption without being explicit about it  undercuts the arguments advanced in other parts of the book about the transparency of the DAG-based assumptions relative to the PO formulation.

A second issue is that TBOW is unhappy with the formulation of the critical assumptions here in terms of the potential outcomes:
\begin{quote} ``To determine if ED [education] is ignorable [...] we are supposed to judge whether employees who would have one potential salary, say $S_1=s$ [that is, $Y({\rm highschool})=s$], are just as likely to have one level of education as the employees who would have a different potential salary, say $S_1=s'$ [that is, $Y({\rm highschool})=s'$]. If you think this sounds circular, I can only agree with you! ... It is quite a cognitive nightmare.''
(TBOW, p. 282).
\end{quote}
These assumptions may be challenging, but of course the formulation agrees closely with standard economic theory. Since \citep*{roy1951some} and \citep*{mincer1974schooling}  economists model individuals choosing whether to go to college partly by comparing their  utility ({\it e.g.,} earnings) if they were to go to college with the utility if they were not to go to college, that is, by comparing their potential outcomes  $Y({\rm highschool})$ and  $Y({\rm college})$, or expecations thereof. It is precisely the important role the potential outcomes play in the formulation of the  choice models ({\it e.g.,} the random utility choice models in  \citep*{mcfadden1973conditional} and \citep*{manski1977structure}) that makes the PO framework a natural one in economics that resonates with practitioners.

\subsection{An Example: Identifying the Returns to Education}

Let me finally discuss an empirical setting where economists have used a variety of identification strategies for the same substantive  problem and compare DAGs and econometric identification strategies. The focus is on estimating the returns to education, or, more specifically, the effect of years of education on the logarithm of earnings. Since \citep*{mincer1974schooling} economists have focused on the effect on the logarithm of earnings in order to interpret the effect as a return, similar to the return on financial investments.

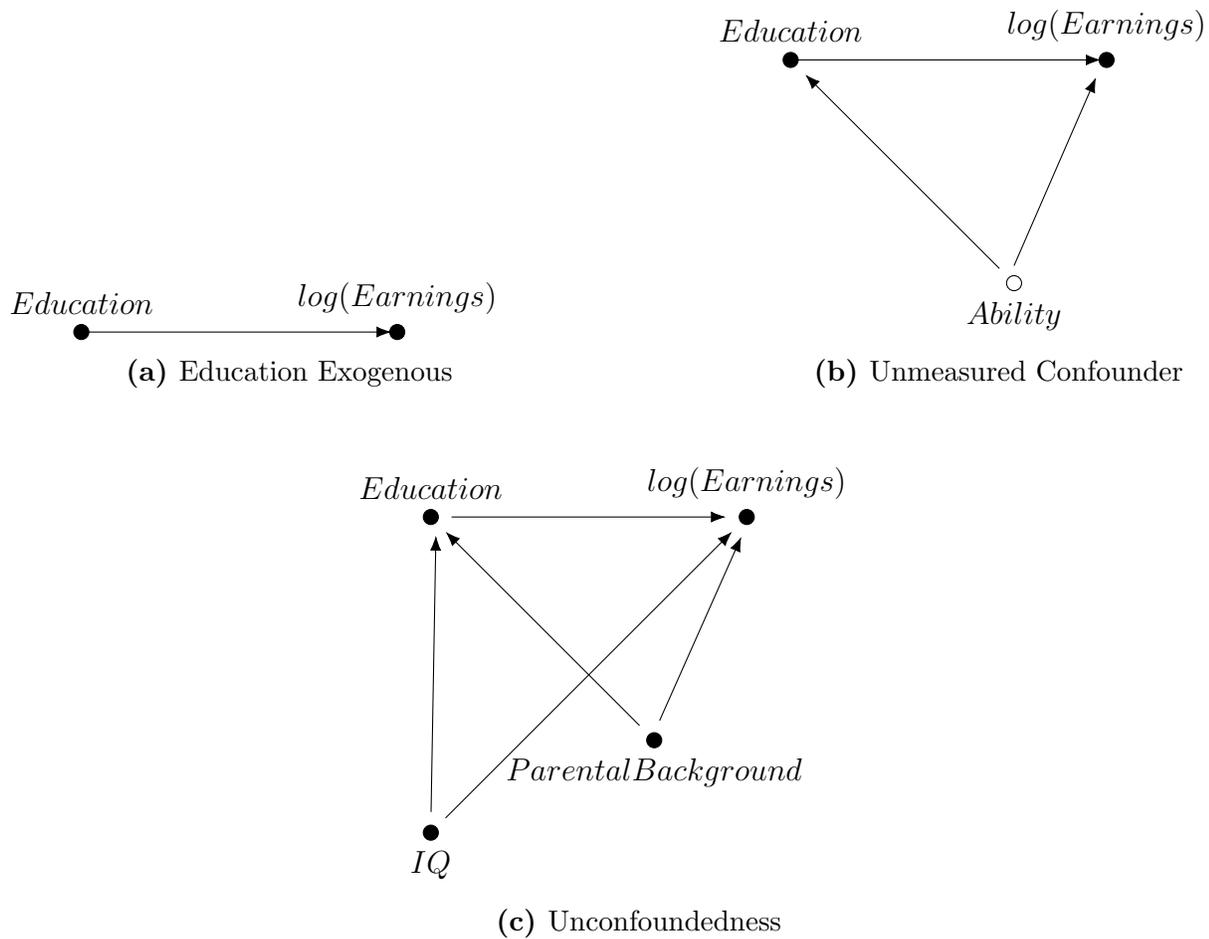
\begin{figure}
    \centering
    \begin{subfigure}[b]{0.45\textwidth}  
    \begin{tikzpicture}[
        >=stealth,
        node distance=4cm        ]
        \node[observed, label=above:{\(Education\)}] (1) {};
        \node[observed, right=of 1,  label=above:{\(log(Earnings)\)}] (2) {};
        \draw [->, notouch] (1.west) -- (2.east);
    \end{tikzpicture}
            \caption{Education Exogenous}
    \end{subfigure}
    \hfill
    \begin{subfigure}[b]{0.45\textwidth}   
    \vspace{2em}
    \begin{tikzpicture}[
        >=stealth,
        node distance=4cm
        ]
        \node[observed, label=above:{\(Education\)}] (1) {};
        \node[observed, right=of 1,  label=above:{\(log(Earnings)\)}] (2) {};
        \node[unobserved, below right=of 1, label=below:{\(Ability\)}] (3) {};
        \draw [->, notouch] (1.west) -- (2.east);
        \draw [->, notouch] (3.north west) -- (1.south east);
        \draw [->, notouch] (3.north west) -- (2.south west);
    \end{tikzpicture}    
        \caption{Unmeasured Confounder}
   
    \end{subfigure}
    \vspace{3em}
    \begin{subfigure}[b]{0.45\textwidth}
        \vspace{2em}
    \begin{tikzpicture}[
        >=stealth,
        node distance=4cm
        ]
        \node[observed, label=above:{\(Education\)}] (1) {};
        \node[observed, right=of 1,  label=above:{\(log(Earnings)\)}] (2) {};
        \node[observed, below right=of 1, label=below:{\(Parental Background\)}] (3) {};
     \node[observed, below=of 1, label=below:{\(IQ\)}] (4) {};
        \draw [->, notouch] (1.east) -- (2.west);
        \draw [->, notouch] (4.north) -- (1.south east);
        \draw [->, notouch] (4.north east) -- (2.south west);
        \draw [->, notouch] (3.north) -- (2.south);
            \draw [->, notouch] (3.north west) -- (1.south east);
    \end{tikzpicture}
        \caption{Unconfoundedness}
    \end{subfigure}
    \hfill

  \caption{\label{fig_education} DAGs for the Returns to Education (I)}
\end{figure}

\begin{figure}
    \centering

\begin{subfigure}
    [b]{0.45\textwidth}\vspace{2em}
    \begin{tikzpicture}[
        >=stealth,
        node distance=2.5cm
        ]
        \node[observed, label=above:{\(Education\)}] (1) {};
        \node[observed, right=of 1,  label=above:{\(log(Earnings)\)}] (2) {};
        \node[unobserved, below right=of 1, label=below:{\(Ability\)}] (3) {};
        \node[observed, left=of 1, label=above:{\(Quarter of Birth\)}] (4) {};
        \draw [->, notouch] (1.east) -- (2.west);
        \draw [dashed, ->, notouch] (3.north west) -- (1.south east);
        \draw [dashed, ->, notouch] (3.north west) -- (2.south west);
        \draw [->, notouch] (4.east) -- (1.west);
    \end{tikzpicture}\vspace{3em}
    \caption{Instrumental Variables: Quarter of Birth}
   \end{subfigure}

\begin{subfigure}
    [b]{0.45\textwidth}\vspace{2em}
    \begin{tikzpicture}[
        >=stealth,
        node distance=3cm
        ]
        \node[observed, label=above:{\(Education\)}] (1) {};
        \node[observed, right=of 1,  label=above:{\(log(Earnings)\)}] (2) {};
        \node[unobserved, below right=of 1, label=below:{\(Ability\)}] (3) {};
        \node[observed, left=of 1, label=above:{\(Distance to College\)}] (4) {};
        \draw [->, notouch] (1.east) -- (2.west);
        \draw [dashed, ->, notouch] (3.north west) -- (1.south east);
        \draw [dashed, ->, notouch] (3.north west) -- (2.south west);
        \draw [->, notouch] (4.east) -- (1.west);
    \end{tikzpicture}\vspace{3em}
    \caption{Instrumental Variables: Distance to College}
   \end{subfigure}

\begin{subfigure}
    [b]{0.45\textwidth}\vspace{2em}
    \begin{tikzpicture}[
        >=stealth,
        node distance=2.5cm
        ]
        \node[observed, label=above:{\(Education\)}] (1) {};
        \node[observed, right=of 1,  label=above:{\(log(Earnings)\)}] (2) {};
        \node[unobserved, below right=of 1, label=right:{\(Genetic Background\)}] (3) {};
        \node[observed, below left=of 3, label=below:{\(Education Twin\)}] (4) {};
        \node[observed, below right=of 3, label=below:{\(log(Earnings) Twin\)}] (5) {};
        \draw [->, notouch] (1.east) -- (2.west);
        \draw [dashed, ->, notouch] (3.north west) -- (1.south east);
        \draw [dashed, ->, notouch] (3.north west) -- (2.south west);
        \draw [dashed, ->, notouch] (3.south east) -- (4.north west);
        \draw [dashed, ->, notouch] (3.south west) -- (5.north east); 
          \draw [->, notouch] (4.east) -- (5.west);
    \end{tikzpicture}\vspace{3em}
    \caption{Fixed Effects Using Twins}
   \end{subfigure}

  \caption{\label{fig_education2} DAGs for the Returns to Education (II)}
\end{figure}
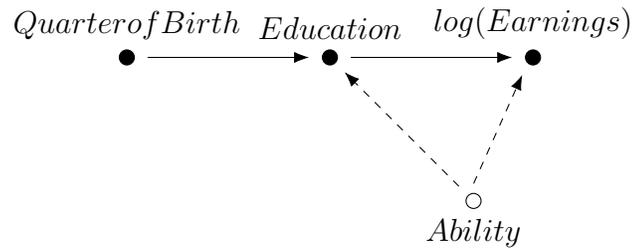
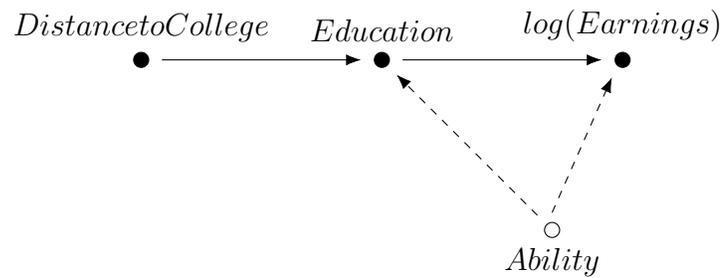
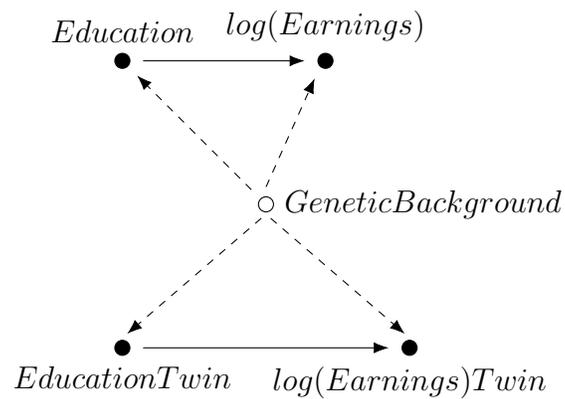

The basic data consists of observations on years of education and the logarithm of earnings. 
Figure \ref{fig_education}(a) shows the simplest strategy in a DAG format, directly comparing people with different levels of education. 
Typically this would be implemented by linear regression of log earnings on years of education, but in a more flexible approach one could simply compare average values for log earnings by education levels.
The concern is that this is too simple a model, and that there may be an unmeasured confounder.
In substantive discussions the unmeasured confounder is typically thought of as unmeasured ability, as in Figure \ref{fig_education}(b). Omitted variable bias arguments suggest that failure to adjust for differences in ability would lead to an upward bias in the estimated returns to education, because it is likely that ability is positively correlated with levels of education and with the potential outcomes.
See \citep*{griliches1977estimating, card1999causal, card2001estimating} for  general discussions. 

In response to these concerns econometricians have used a variety of approaches. I will briefly discuss some of them and attempt to illustrate them in DAGs.
One approach is to attempt to measure ability. For example, one may use parental background  measures, including parental education or socio-economic status.
Researchers have also attempted to directly measure ability, through test scores or iq tests. Using both parental background and iq scores leads to a DAG  as in Figure  \ref{fig_education}(c) where adjusting for the pre-treatment variables leads to unbiased estimates of the effect of interest.
Of course there may be concerns that variables like parental background do not have a direct causal effect on the child's later earnings, and that this is merely a proxy for unobserved ability. In practice, however, researchers have attempted to control as much as possible for variables that are proper pre-treatment (that is, prior to educational decisions being made), without concerns about introducing M-bias. It is rare to see researchers use post-treatment variables as controls in this approach.

Another strategy has been to use instrumental variables approaches, based on instruments that change the incentives to acquire additional formal education. One classic example is the 
\citep*{angristkrueger1991} study using quarter of birth as an instrument, as in Figures \ref{fig_education2}(a). A second classic example is \citep*{card1993using} using distance to college as an instrument, as in
 \ref{fig_education2}(b). 
Both applications are complicated. Although originally they assumed constant treatment effects, they are now typically interpreted as  weighted average effects for compliers with the weights varying by education levels \citep*{angrist1995two}. There are also concerns with the exclusion restriction in both cases. There may also be concerns with the monotonicity condition in the college distance application if the closest college is a two-year college. The \citep*{angristkrueger1991} study also spawned a big literature on weak instrument problems ({\it e.g.,} \citep*{stock1997, andrews2007, chamberlain}). These problems are often naturally discussed in a PO framework, with the benefit of putting this in a DAG unclear.

Yet another strategy has been to look for siblings, or, going even further, twins, as in \citep*{ashenfelter1994estimates, ashenfelter1998income}. The estimation procedure regresses differences in the logarithm of earnings on differences in years of education. Implicitly there is a DAG similar to Figures \ref{fig_education2}(c), but there is additional structure.
The causal effect of education on log earnings is the same for the two members of a twin pair, and similarly the causal effect of the unobserved confounder (genetic background) on education and earnings is the same for both members of the twin. The full set of assumptions here is difficult to capture in a DAG.
There is some connection between this identification strategy and the multiple causes approach in \citep*{wang2018blessings}.  They assume that there is an unobserved confounder that affects both causes. In the twins case this is the genetic background that is common to both members of the twin pair. The additional structure in the twins case (symmetry between the two members of the twin pairs) helps pin down the shared confounder, but there may be additional links that can be exploited. These strategies are not without controversy: it is not always clear that the variation in education levels between siblings or twins is exogenous (see \citep*{card1999causal, card2001estimating}).

This is a rich literature full of clever attempts to get closer to the causal effect of education. The challenge for the graphical literature is to come up with novel strategies that shed light on substantive questions like this.

\section{Conclusion}

TBOW and the DAG approach fully deserve the attention of all researchers and users of causal inference as one of its leading methodologies. Is it more than that? Should it be the framework of choice for all causal questions, everywhere, or at least  in the social sciences, as TBOW argues? 
Should it be the starting point for teaching about causality?
In my view  the answer to both questions is no. The questions it currently answers well are not the ones that are the most pressing ones in practice. Conversely, for the most common  and important questions the PO framework is in my view an attractive one in social sciences, and one that resonates well with economic theory by effectively incorporating restrictions beyond conditional independencies. Moreover, for the problems where DAGs could contribute substantially, the most important issue holding back the DAGs is the lack of convincing empirical applications. History suggests that credible applications are what drives the adoption of new methodologies in economics and other social sciences, not  mathematical elegance or rhetoric. It is studies such as
\citep*{lalonde}, \citep*{cardmariel}, \citep*{angrist1990lifetime},  \citep*{angristkrueger1991}, and \citep*{ashenfelter1994estimates} that spurred the credibility revolution and the adoption of the PO framework, not the theoretical advances  -- they came later. There have not been similar applications of the DAG framework, and more papers discussing toy models will not be sufficient to convince economists to use this framework.

\clearpage

\bibliographystyle{plainnat}
\bibliography{references}

\end{document}